\documentclass[fleqn,usenatbib]{mnras}

\usepackage{newtxtext,newtxmath}

\usepackage[T1]{fontenc}
\usepackage{graphicx}
\usepackage{subcaption}
\usepackage{adjustbox}
\usepackage{multicol}
\usepackage{paralist}

\DeclareRobustCommand{\VAN}[3]{#2}
\let\VANthebibliography\thebibliography
\def\thebibliography{\DeclareRobustCommand{\VAN}[3]{##3}\VANthebibliography}

\usepackage{graphicx}	
\usepackage{amsmath}	

\title[Binary planet formation through tides]{Binary planet formation through tides}

\author[C. Lazzoni et al.]{
C. Lazzoni,$^{1}$\thanks{E-mail: c.lazzoni@exeter.ac.uk}
K. Rice,$^{2,3}$
A. Zurlo,$^{4,5,6}$
S. Hinkley,$^{1}$
S. Desidera$^{7}$
\\
$^{1}$University of Exeter, Physics Building, Stocker Road, Exeter, EX4 4QL, UK\\
$^{2}$SUPA, Institute for Astronomy, Royal Observatory, University of Edinburgh, Blackford Hill, Edinburgh EH93HJ, UK\\
$^{3}$Centre for Exoplanet Science, University of Edinburgh, Edinburgh, UK\\
$^{4}$Instituto de Estudios Astrofísicos, Facultad de Ingeniería y Ciencias, Universidad Diego Portales, Av. Ejercito 441, Santiago, Chile\\
$^{5}$Escuela de Ingeniería Industrial, Facultad de Ingeniería y Ciencias, Universidad Diego Portales, Av. Ejercito 441, Santiago, Chile\\
$^{6}$Millennium Nucleus on Young Exoplanets and their Moons (YEMS), Santiago, Chile\\
$^{7}$Osservatorio Astronomico di Padova, INAF, Vicolo
dell’Osservatorio 5, Padova, Italia.
}

\date{Accepted XXX. Received YYY; in original form ZZZ}

\pubyear{2023}

\begin{document}
\label{firstpage}
\pagerange{\pageref{firstpage}--\pageref{lastpage}}
\maketitle

\begin{abstract}
The search for satellites around exoplanets represents one of the greatest challenges in advancing the characterization of planetary systems. Currently, we can only detect massive satellites, which resemble additional planetary companions rather than rocky moons. It is not yet well understood whether such substellar pairs, known as binary planets, are common or how they form. In this study, we investigated the formation scenario for binary planets resulting from tidal dissipation during close encounters in the gravitational instability scenario. We conducted seven sets of simulations, varying the number of initial planets injected into the system from two to five, as well as the amount of energy lost due to tides. Our results demonstrate that this formation mechanism is quite efficient in producing binary planets, with an average occurrence rate for the simulated systems of $14.3\%$. Additionally, we present the distribution of relevant physical parameters (semi-major axis, eccentricity, mass ratios, and formation time) for planet-planet pairs. We also provide comprehensive statistics for single planets and planet-planet pairs. \\

\end{abstract}




\section{Introduction}
Planet-planet pairs, also known as binary planets, are among the most enigmatic and captivating objects in the field of exoplanets. Although there is no precise definition for this type of system, here we refer to binary planets as two celestial bodies of similar size that are gravitationally bound and orbit a common center of mass.

The closest example we have, both in terms of proximity and adherence to the definition, is the Pluto-Charon dwarf planet system. Pluto, with a radius of approximately $\sim 1100$ km, is the larger of the two bodies, while Charon has roughly half the radius of Pluto, around $580$ km \citep{Reinsch}. With a mass ratio of $\sim0.12$, Pluto and Charon orbit their common center of mass, which lies outside the surface of Pluto. Over time, they have become gravitationally locked, exhibiting synchronous rotation due to the gravitational forces between them \citep{Grundy}.

Several free-floating binary sub-stellar objects have already been detected, although statistical analysis indicates that they are rare \citep{fontanive2023}. Although more massive than the Pluto-Charon system and often surpassing the deuterium burning limit, these systems exhibit mass ratios in the range $[0.1,1]$ Examples at the lower end of this mass range include 2MASS J11193254–1137466 AB \citep{Best}, where the primary has a mass of approximately 25 M\textsubscript{Jup} and the secondary has a mass of $\sim 14$ M\textsubscript{Jup} (mass ratio of 0.56). Another example is Oph 11 Ab \citep{Allers,Close} and 2MASSW J1207334-393254 Ab \citep{Song}, where the primaries are in the brown dwarf regime, with masses of approximately 18 M\textsubscript{Jup} and 30 M\textsubscript{Jup} respectively, while the secondaries are planets with masses of 10 M\textsubscript{Jup} and 5 M\textsubscript{Jup} respectively (mass ratios of 0.55 and 0.2).

Very recently, \citet{Pearson} identified several free-floating objects in the inner Orion Nebula and Trapezium Cluster using JWST. Unexpectedly, they found that $9\%$ of the planetary mass objects are in wide binaries
and they have binary fraction larger than that of brown dwarfs in the same environment and in the same range of separations.

While stars hosting massive brown dwarf pairs with mass ratios close to 1 have been observed (e.g. $\epsilon$ Indi, \cite{King}; GJ 569 \cite{Femenia}; GJ417 \cite{Kirkpatrick}), the confirmed detection of a planet-planet pair orbiting a central star is still pending. Recently, a few candidates have been proposed, such as the Kepler 1625 b-i \citep{Teachey}, Kepler 1708 b-i \citep{Kipping}, and DH Tau Bb \cite{Lazzoni2} systems. The identification of the Kepler 1625 b-i and Kepler 1708 b-i systems is based on the analysis of transit timing variations (TTVs) and shallow dips in the light curves of the stars, using mostly Kepler and, for the first target, Hubble Space Telescope (HST) observations. On the other hand, the DH Tau Bb system was detected through direct imaging.

There are multiple mechanisms that can explain the formation of binary planets, including giant impacts \citep[limited to rocky bodies][]{Goldreich,Woolfson,Koch}, formation within circumplanetary disks \citep{Inderbitzi2020}, and the prompt formation of binaries from circumstellar disks, similarly to stellar pairs \citep[see e.g.][]{Kuruwita}. In this study, our focus is on gravitational capture enhanced by tidal effects. In this scenario, the two planets are independently formed within the circumstellar disk and subsequently become a bound pair due to close encounters, during which energy is dissipated through tidal interactions. Our study was inspired by \cite{Ochiai}, where the authors investigated the formation of binary planets through gravitational capture, assuming a planetary population characterized by core accretion models \citep[CA, ][]{Mizuno}. Their main finding was that approximately one system out of ten could host a planet-planet pair.

While the studies conducted by Ochiai could explain the existence of pairs such as Kepler 1625 b-i, sub-stellar objects like DH Tau Bb are unlikely to have formed through core accretion, given their masses (10-20 M\textsubscript{Jup}) and large separations (hundreds of AU). Instead of core accretion, which can easily explain the formation of companions ranging from small terrestrial planets to Jupiter-like objects within a few astronomical units from the star, the gravitational instability scenario \citep[GI,][]{Cameron, Boss} is better suited to explain the presence of massive companions ($\gtrsim$ 5 M\textsubscript{Jup}) located farther out in the system.\\
In the following sections, we will present the results of dynamical simulations aimed at reproducing the formation of planetary pairs through orbital crossing and tidal dissipation. These simulations start with a population of gravitational instability-like planets.

Section 2 describes the modifications made to the Mercury code for conducting the simulations, as well as the specific settings used. In Section 3, we present the results related to the formation of binary planets, along with statistics on both bound and free-floating single and binary substellar objects. Furthermore, in Section 4, we compare our results with information obtained from observational surveys. Finally, in Section 5, we discuss our conclusions based on the findings of this study.

\section{Simulations}
To simulate the formation of binary planets we used a modified version of the Mercury Code \citep{Chambers}, adding a term that accounts for energy dispersion during close encounters. The energy dissipation factor comes from tidal interaction between the planets and was taken from \cite{Ochiai}.

When two planets have a close encounter, they can lose some orbital energy through tidal interactions.  For a single close encounter between two planets, $i$, and $j$, the energy loss due to this tidal interaction is
\begin{equation}
\begin{split}
E_{\rm tide} = \frac{1}{\lambda} \Biggl(\frac{G M_j}{R_i} \left[\left( \frac{R_i}{q_{ij}}\right)^6 T_2(\eta_i) +
\left( \frac{R_i}{q_{ij}}\right)^8 T_3(\eta_i) \right] + \\ \frac{G M_i}{R_j} \left[\left( \frac{R_j}{q_{ij}}\right)^6 T_2(\eta_j) + \left( \frac{R_j}{q_{ij}}\right)^8 T_3(\eta_j) \right] \Biggr),
\label{ochiai}
\end{split}
\end{equation}
where $M_i$, $M_j$, $R_i$ and $R_j$ are the masses and radii of the planets, $q_{ij}$ is the pericentre distance between the planets, $\eta_{i}~\equiv~[M_{i}/(M_i + M_j)]^{1/2}(q_{ij}/R_i)^{3/2}$, and $T_2$ and $T_3$ are fifth-degree polynomials, with coefficients given in \citet{Portegies}.  We use the coefficients for an $n = 2$ polytrope and assume that both planets have radii of 1.5 R\textsubscript{Jup}. Finally $\lambda = 1, 10, 100$ is the term considered to modulate the strength of tides.

We implement the above in the same way as \citet{Ochiai}. If two planets have a close encounter, then at pericenter passage we first move into the centre of mass frame of the two planets.  We then subtract the energy dissipated by tides, given by Equation \eqref{ochiai}, from the energy of the two-planet system, keeping the direction of the velocity vectors of the two planets unchanged and also ensuring that the velocity of the centre of mass of the two planet system doesn't change. 

We note that this approximation is valid as long as the time between encounters is long enough that the energy has been dissipated in the planets. This process is, in fact, only able to describe the first stages of planetary interactions. The impulse approximation and the use of Equation (\ref{ochiai}) is no longer really appropriate once the two planets are in a bound orbit and the orbital eccentricity has been significantly damped.  However, the goal of this work is to consider the efficiency of gravitational capture enhanced by tides in forming binary planets, rather than trying to constrain the intrinsic characteristics of the binary such as the mutual eccentricities and separation between the two planets.  Hence, we stopped the integration and labeled any system as a binary when the two planets reach both a pericentre distance and semi-major axis of less than 0.1 au ($\sim$210 R\textsubscript{Jup}). We note that this is above the Roche limit for tidal disruption which, at most, can reach $\sim$3 R\textsubscript{Jup}.

Over time, we might expect tidal effects to continue to shrink the orbit of the binary planets system, eventually tending towards a synchronous state \citep{Ochiai}.  However, this is likely to be very slow and should not be relevant for the relatively young systems that we are considering here.  Moreover, since the binaries formed are usually on wide orbits, the influence of the parent star should be weak enough to allow the survival of the latter \citep{Sasaki12}.  

For the simulations that were stopped due to the formation of a planet-planet pair, we resumed the integration considering the binary as a single object with a mass given by the sum of the masses of the planets and an initial position and velocity the same as that of the center of mass of the binary system. In a few cases, triple/quadruple systems were formed. However, the simplistic assumptions used to resume the integration are not suitable to describe the dynamics for more than two bound planets. We therefore discarded such outcomes from our analysis. However, these results might point toward the existence of planetary configurations with multiplicity higher than two.

We considered seven different set-ups: two, three, four, and five initial planets with $\lambda=1$, and a random number of planets taken from the range $[2,5]$ with $\lambda=1$, $\lambda=10$ and $\lambda=100$. For every set-up, we run ten sets of simulations each with 100 systems for a total time of 1.5Myr. We used the Bulirsch-Stoer \citep[BS][]{Bulirsch} integrator, which can account for non-conservative systems, with time-steps of 0.5 days and an accuracy parameter of $10^{-10}$. Masses and semi-major axes for the planets injected in each system were randomly selected in the ranges [1,15] M\textsubscript{Jup} and [50,100] au to roughly reflect the typical parameters distribution of planets formed via GI \citep[e.g.,][]{Forgan}. 

The initial eccentricity for each planet was randomly selected in the range $[0,0.1]$, inclination in the range $[0^{\circ},0.1^{\circ}]$, and longitude of periapsis, mean anomaly and argument of periapsis in the range $[0^{\circ},360^{\circ}]$.

Since we expect that the planets are still contracting, we accounted for an inflated radius of 1.5 R\textsubscript{Jup}, independently of the mass. This is justified by the fact that giant planets' radii usually show a weak dependence on mass \citep[$R\propto M^{0.01}$ for $M>0.39$ M\textsubscript{Jup},][]{Bashi}.
The central star has fixed parameters of $1 M_{\odot}$ and $R=0.005$ au and the ejection distance for a planet was set at $10^4$ au. A summary of the simulations setup is given in Table \ref{tab1}.

\begin{table}
	\centering
	\caption{Initial parameters for the seven set-ups used for the integrations. From left to right, identification number of each set, number of systems simulated, number of initial planets, and adjustment factor for energy lost due to tides (see Equation \eqref{ochiai}).}
	\label{tab1}
	\begin{tabular}{cccr} 
		\hline
		Set-up & Systems & N & $\lambda$ \\
		\hline
		1 & $1000$ & 2 & 1\\
		2 & $1000$ & 3 & 1\\
		3 & $1000$ & 4 & 1\\
            4 & $1000$ & 5 & 1\\
            5 & $1000$ & random(2,5) & 1\\
            6 & $1000$ & random(2,5) & 10\\
            7 & $1000$ & random(2,5) & 100\\
		\hline
	\end{tabular}
\end{table}

\section{Results}
Table \ref{tab2} provides a summary of the results obtained from the seven sets of simulations. 

For the first four sets, a total of thirty-five systems had to be discarded due to the presence of triple/quadruple planetary systems (twelve systems for set 2, eleven for set 3, and twelve for set 4).  

Moreover, for the last three sets where energy loss varied according to $1/\lambda$, a total of twenty-seven systems were discarded (twelve for set 5, seven for set 6, and eight for set 7).

The percentage of discarded systems due to the formation of configurations with more than two bound planets is shown in the last column of Table \ref{tab2}. 

The first three columns of Table \ref{tab2} provide information on the percentage of systems that formed binary planets, the binaries that were ejected, and those that collided with the central star. Notably, some of the systems were able to form two pairs of binary planets each (four in set 4, four in set 5, four in set 6, and one in set 7) and were counted twice in the relevant statistics. It is important to note that the first sub-column, dedicated to the rate of systems that formed binary planets, includes both binaries that survived in a stable orbit around the star throughout the integration and those that were either ejected or collided.

Column five displays the percentage of planets that were still orbiting their host star with a semi-major axis $< 10000$ au after 1.5 Myrs. Column six lists the percentage of planets that were ejected, while column seven accounts for the percentage of planets that collided with the central star. These three columns include the counts for both binary planets that survived and those that were ejected or collided. We note that each binary (formed, ejected, or collided) was counted as two individual planets for these statistics.

\begin{table*}
	\centering
	\caption{Results for the seven sets of simulation. In the columns, from left to right: the identification number of the set, the percentage of systems that formed binary planets, the percentage of systems with ejected and collided binaries, the percentage of planets on stable orbits, ejected planets and collided planets, and the percentage of systems that were discarded from the analysis. For these last three columns binaries were included and counted as two planets.}
	\label{tab2}
	\begin{tabular}{cccccccc} 
		\hline
		\multicolumn{1}{c}{Set} & \multicolumn{3}{c}{Binaries} & \multicolumn{3}{c}{Planets}  & \multicolumn{1}{c}{Discarded systems ($\%$)} \\
     &  Formed ($\%$) & Ejected ($\%$) & Collided ($\%$) & Survived ($\%$)&Ejected ($\%$)&Collided ($\%$)& \\
		\hline
		1     & 15.1    &     0  & 0      &  75.6  &   24   & 0.4      &       0\\
            2     & 10.1    &    0.5 & 0.1    &  63.6  &   35.1 & 1.3      &     1.2\\
            3     & 13.5    &    1.7 & 0      &  56.2  &   41.7 & 2.1      &     1.1\\
            4     & 18.6    &    2.6 & 0      &  51.6  &   45.8 & 2.6      &     1.2\\
            5     & 17.2    &   2    &  0     &  58.7  &  39.3  &  2       &    1.2 \\
            6     & 15.7    &  1.2   &  0.1   &  59.8  &  38.3  &  1.8     &     0.7\\
            7     & 14.4    &  1.6   &  0     &  58.7  &  39.3  &  2       &     0.8\\

		\hline
	\end{tabular}
\end{table*}

\subsection{Parameters dependency on the number of initial planets}

In the first four sets of simulations, we varied the initial number of planets, ranging from a minimum of two to a maximum of five.

\begin{figure}
\centering
\includegraphics[width=\columnwidth]{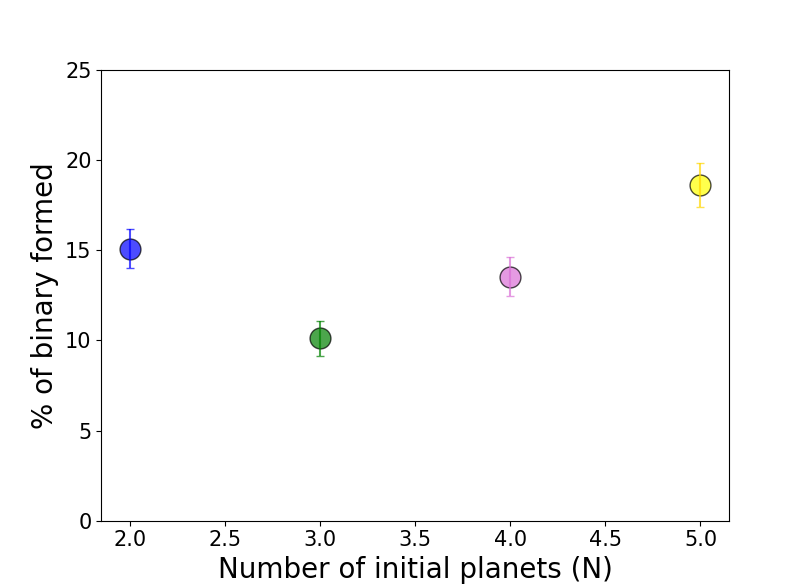} 
\caption{Percentage of systems that formed binary planets with respect to the initial number of planets injected for the first four sets of simulations.}
\label{fig1}
\end{figure}

In Figure \ref{fig1}, we show the percentage of binary planets formed with respect to $N$. The error bars are calculated considering our simulations as a Bernoulli trial where the outcomes can be either success or failure in forming a binary. As depicted by Figure \ref{fig1}, the formation of binary planets is influenced by the initial number $N$ of injected planets. Excluding the first case with $N=2$ for which the formation rate is $15\% \pm 1.1\%$, there is a gradual increase of the latter with $N$. In fact, while the formation rate for $N=3$ is approximately $10\% \pm 0.96\%$, it can be enhanced by up to $8.5\%  \pm 1.2\%$ when considering five planets in the system. The anomaly given by the case with $N=2$ is likely due to the small degree of dynamical scattering and perturbations in the system which enhances the formation of bound pairs. 

It is important to note that the reported percentages encompass all the binaries formed, without distinguishing between binaries that remain in a stable orbit around the central star until the end of the simulation, and those that were ejected from the system or collided with the star.

\begin{figure*}

\centering

\begin{tabular}{@{} c c @{}}
  \begin{tabular}{@{} c c @{}}
    \includegraphics[width=5cm,height=4cm]{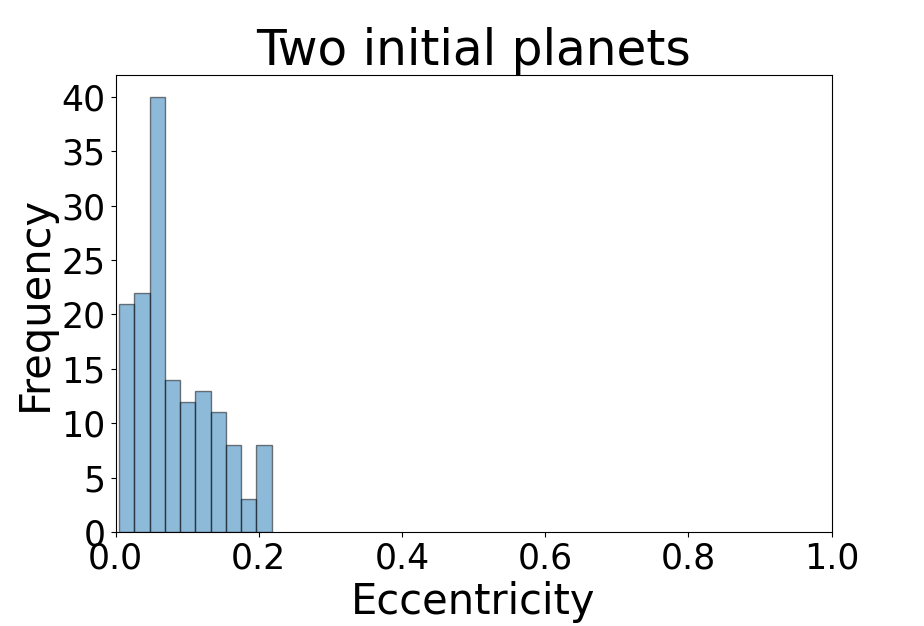} &
    \includegraphics[width=5cm,height=4cm]{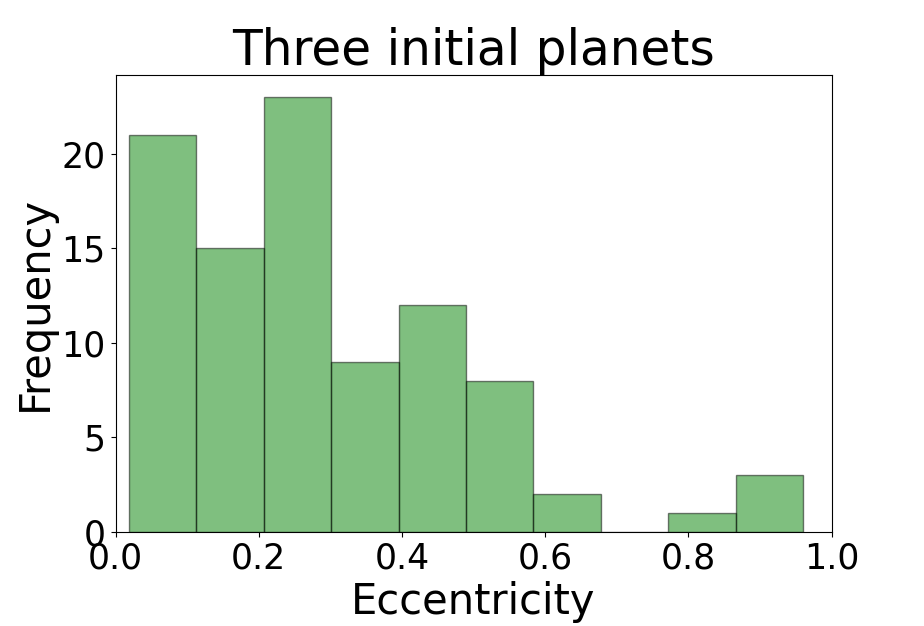} \\[2ex]
    \includegraphics[width=5cm,height=4cm]{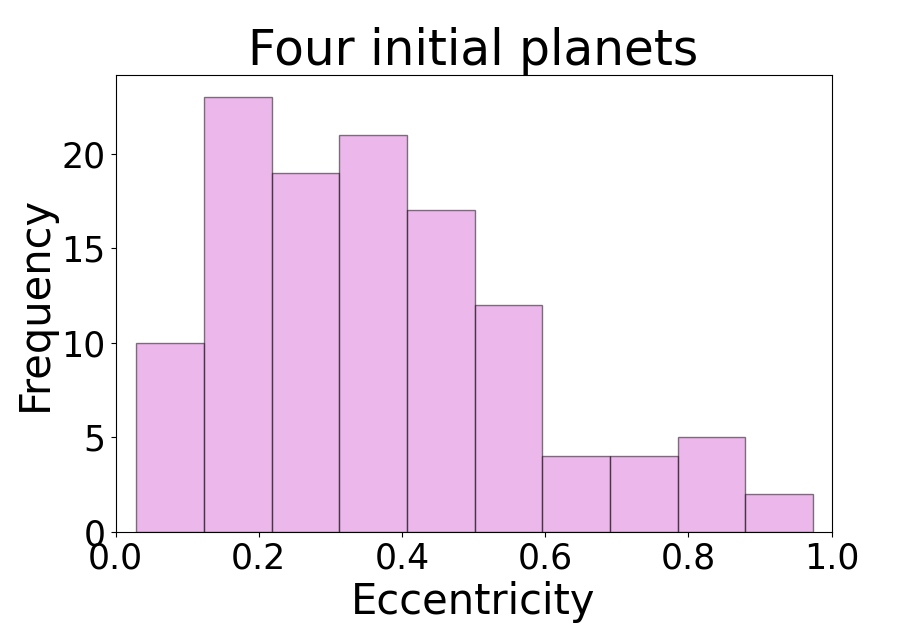} &
    \includegraphics[width=5cm,height=4cm]{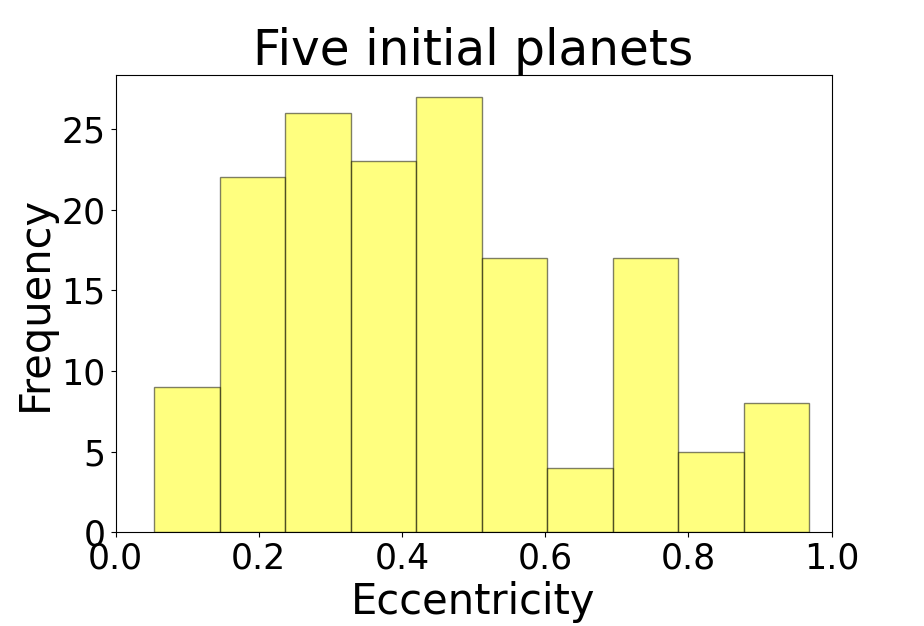}
  \end{tabular}
&
  \begin{tabular}{@{} c @{}}
    \includegraphics[width=7cm,height=6cm]{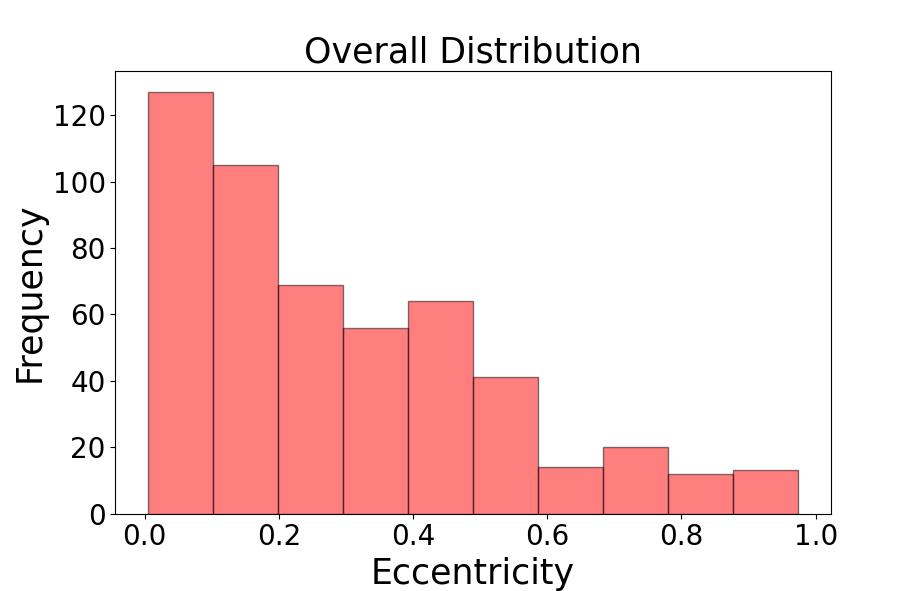}
  \end{tabular}
\end{tabular}

\caption{Eccentricity distributions for binary planets on stable orbits around the star. The eccentricity is that of the orbit of the binary, considered as one object, around the star. On the left, the results from simulations with two (blue), three (green), four (pink), and five (yellow) initial planets. On the right, overall distribution obtained by combining the previous data sets.}
\label{fig2}
\end{figure*}

\begin{figure*}

\centering

\begin{tabular}{@{} c c @{}}
  \begin{tabular}{@{} c c @{}}
    \includegraphics[width=5cm,height=4cm]{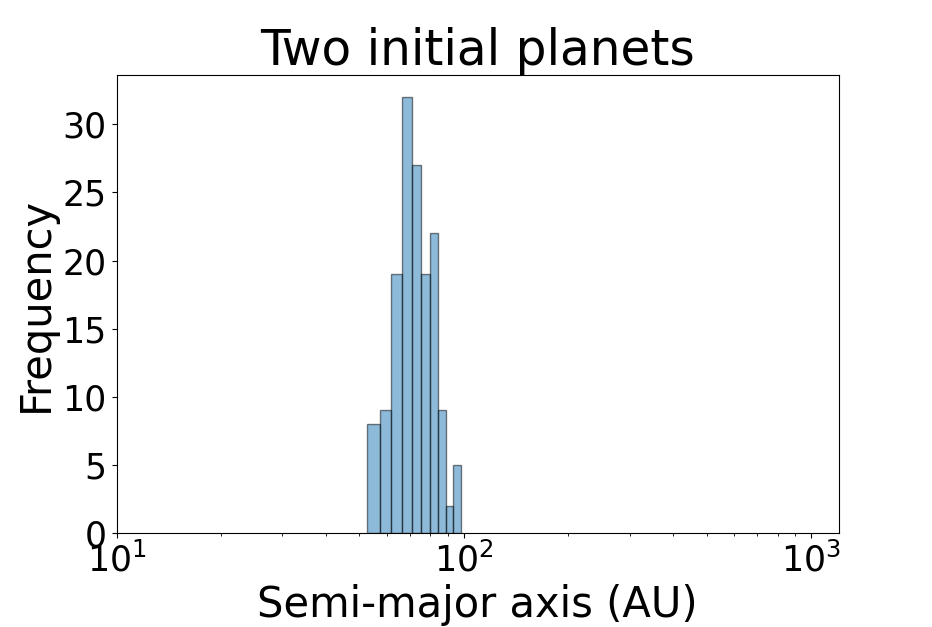} &
    \includegraphics[width=5cm,height=4cm]{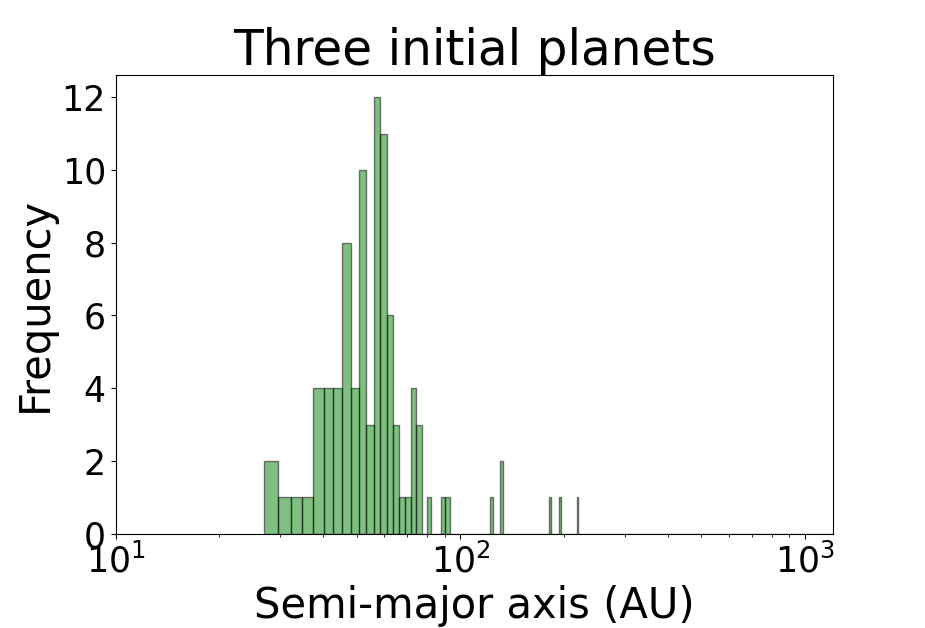} \\[2ex]
    \includegraphics[width=5cm,height=4cm]{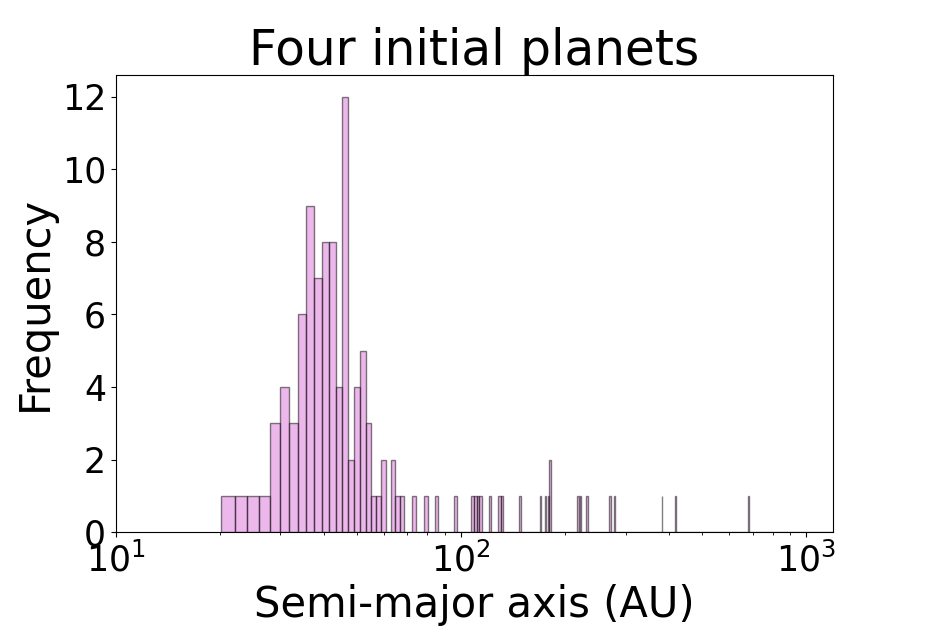} &
    \includegraphics[width=5cm,height=4cm]{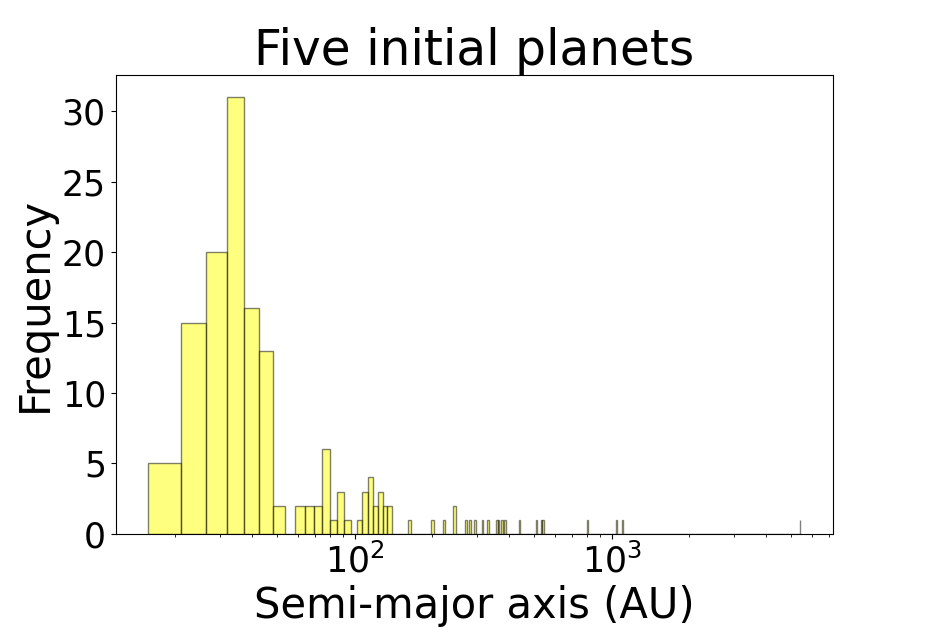}
  \end{tabular}
&
  \begin{tabular}{@{} c @{}}
    \includegraphics[width=7cm,height=6cm]{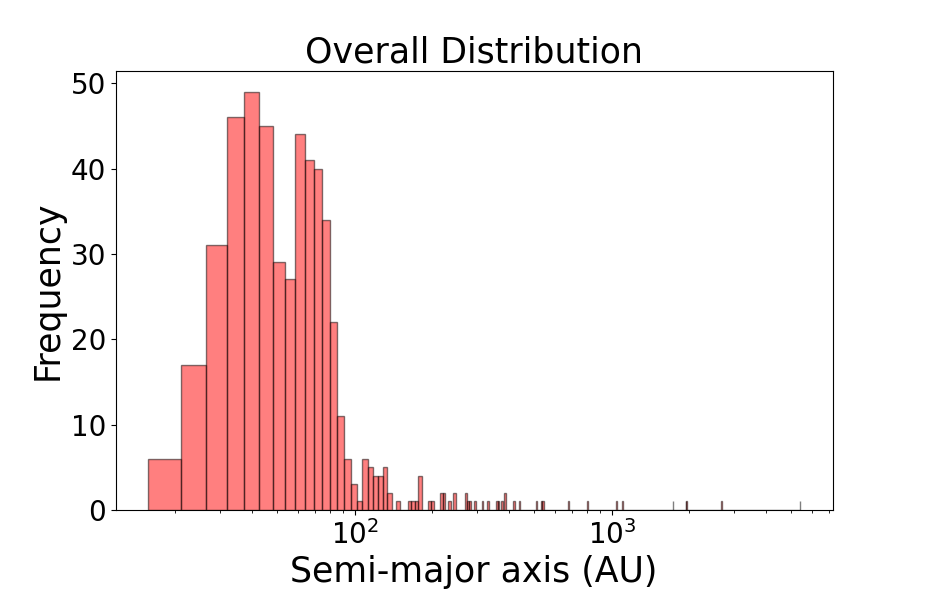}
  \end{tabular}
\end{tabular}

\caption{Semi-major axis distributions for binary planets on stable orbits around the star. The semi-major axis is relative to the orbit of the binary, considered as one object, around the star On the left, results from simulations with two (blue), three (green), four (pink), and five (yellow) initial planets. On the right, overall distribution obtained by combining the previous data sets.}
\label{fig3}
\end{figure*}

\begin{figure*}

\centering

\begin{tabular}{@{} c c @{}}
  \begin{tabular}{@{} c c @{}}
    \includegraphics[width=5cm,height=4cm]{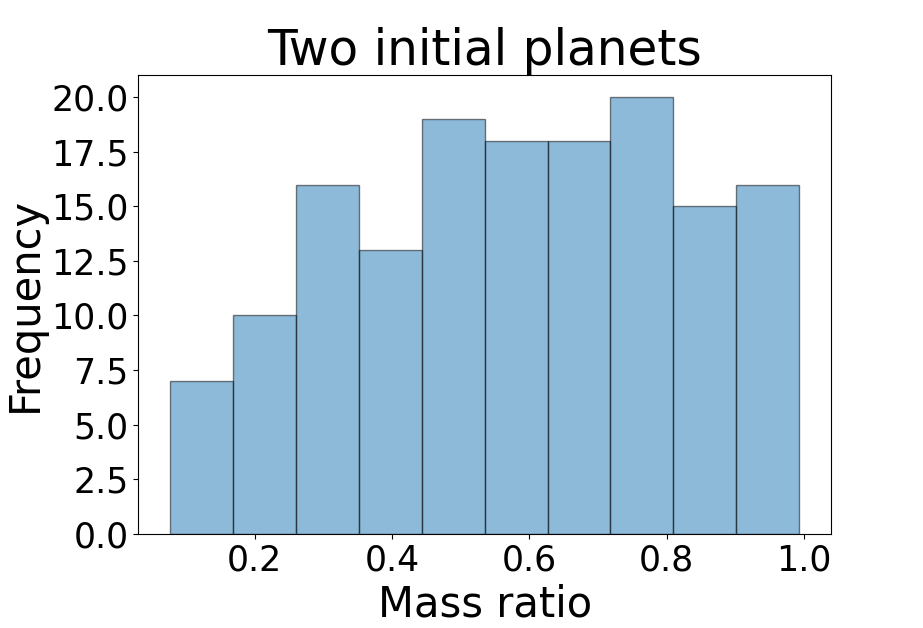} &
    \includegraphics[width=5cm,height=4cm]{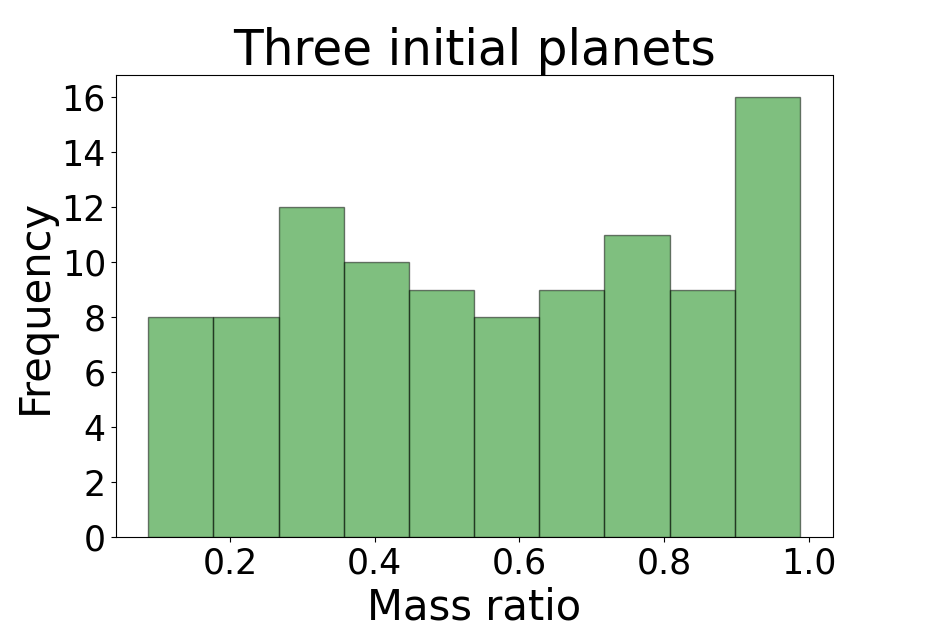} \\[2ex]
    \includegraphics[width=5cm,height=4cm]{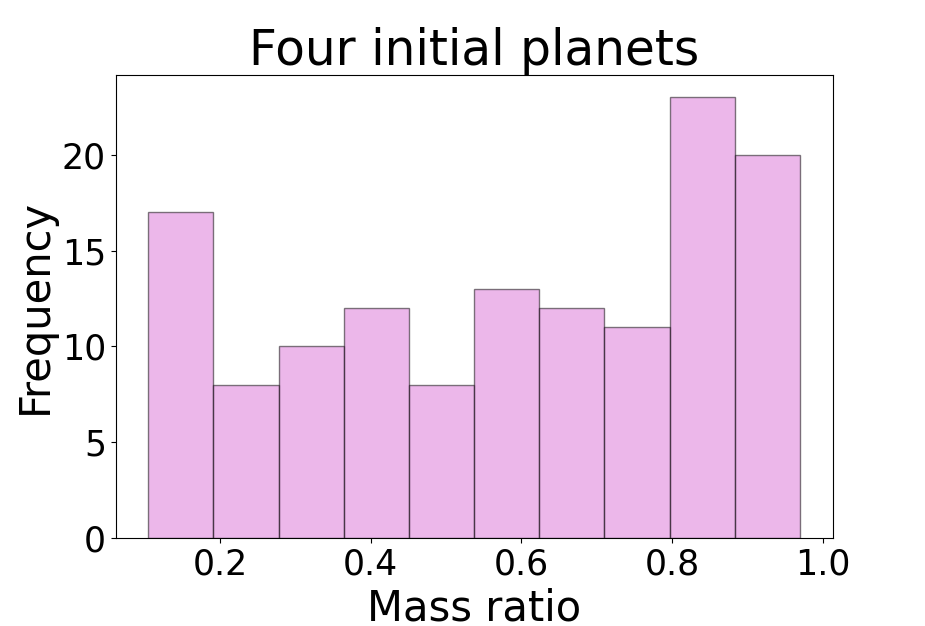} &
    \includegraphics[width=5cm,height=4cm]{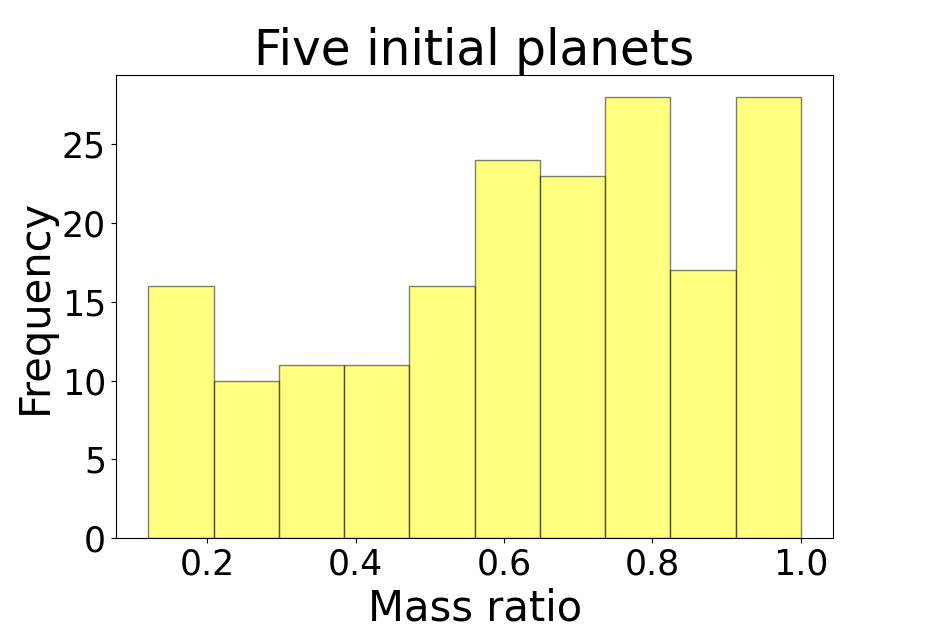}
  \end{tabular}
&
  \begin{tabular}{@{} c @{}}
    \includegraphics[width=7cm,height=6cm]{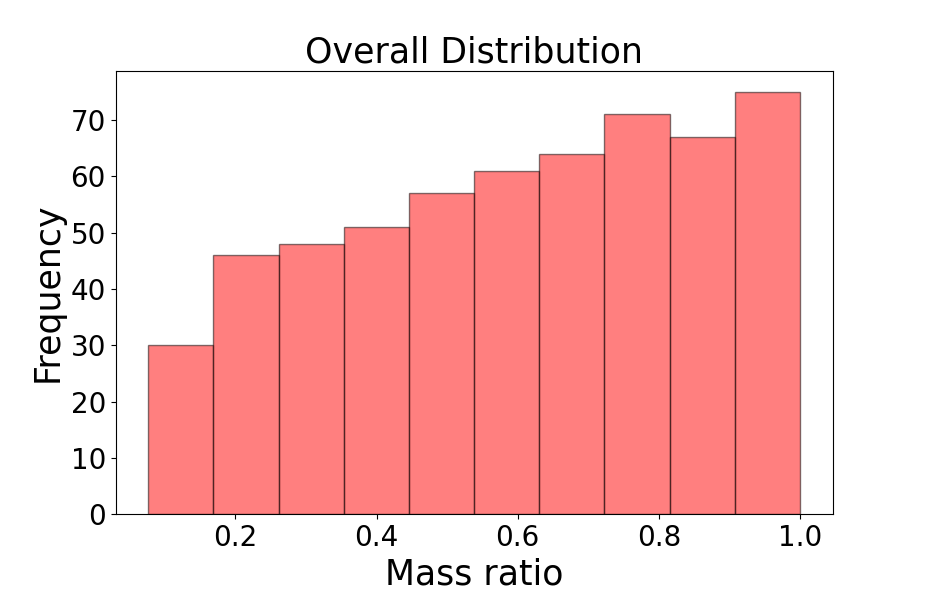}
  \end{tabular}
\end{tabular}

\caption{Mass ratio distributions for every binary planet formed. On the left, results from simulations with two (blue), three (green), four (pink), and five (yellow) initial planets. On the right, overall distribution obtained by combining the previous data sets.}
\label{fig4}
\end{figure*}

\begin{figure*}

\centering

\begin{tabular}{@{} c c @{}}
  \begin{tabular}{@{} c c @{}}
    \includegraphics[width=5cm,height=4cm]{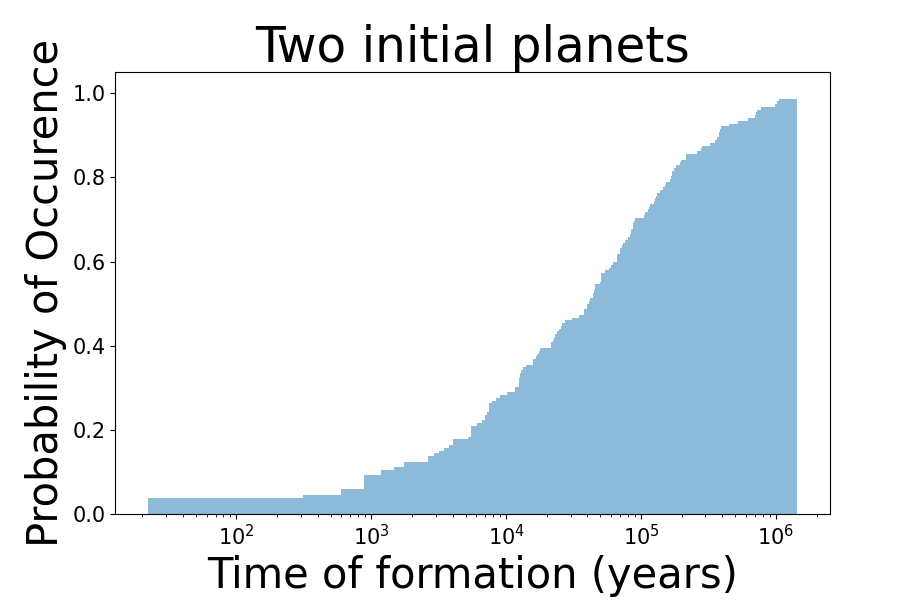} &
    \includegraphics[width=5cm,height=4cm]{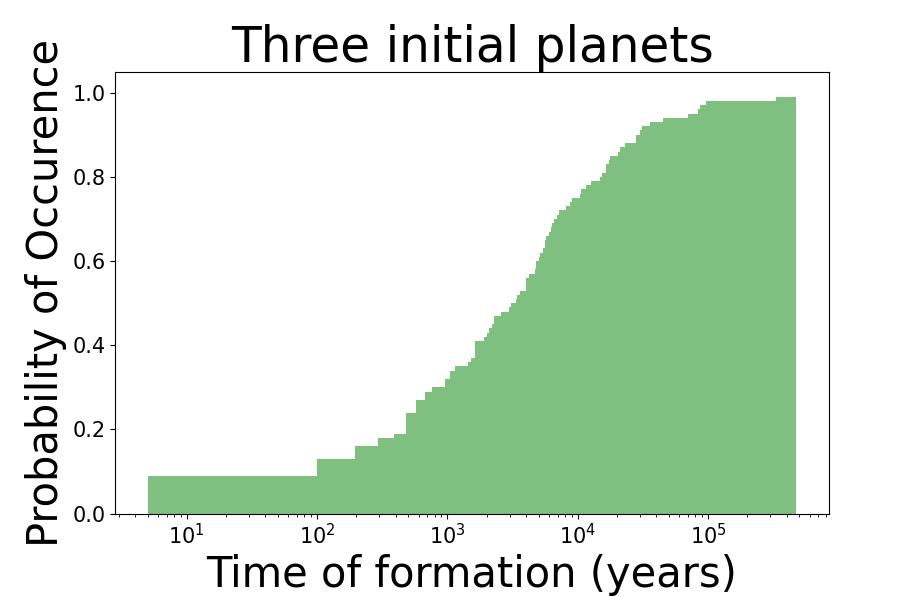} \\[2ex]
    \includegraphics[width=5cm,height=4cm]{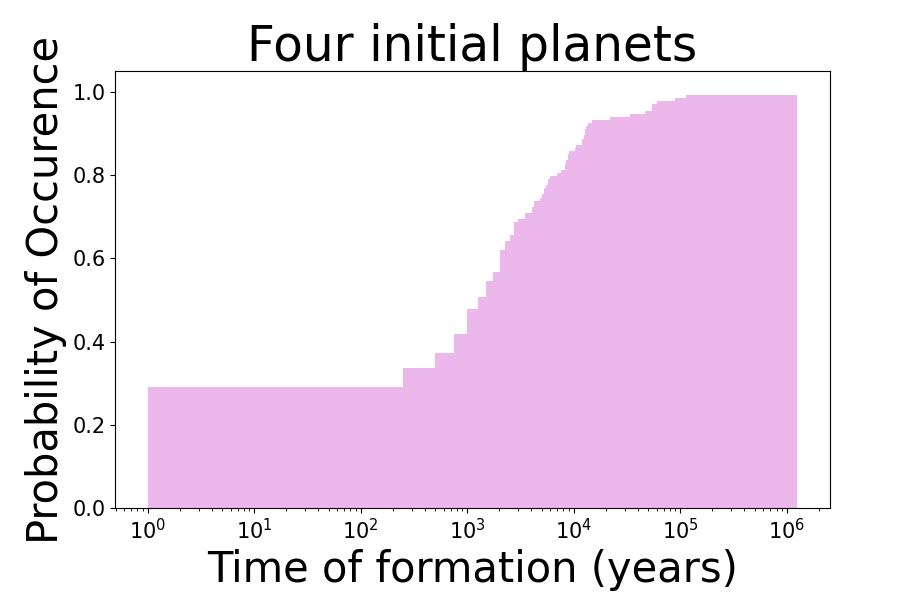} &
    \includegraphics[width=5cm,height=4cm]{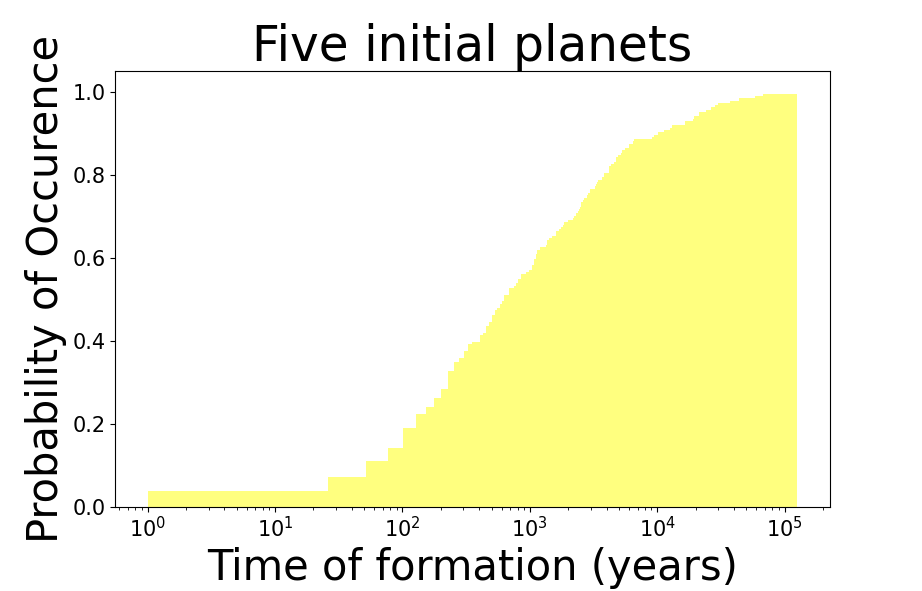}
  \end{tabular}
&
  \begin{tabular}{@{} c @{}}
    \includegraphics[width=7cm,height=6cm]{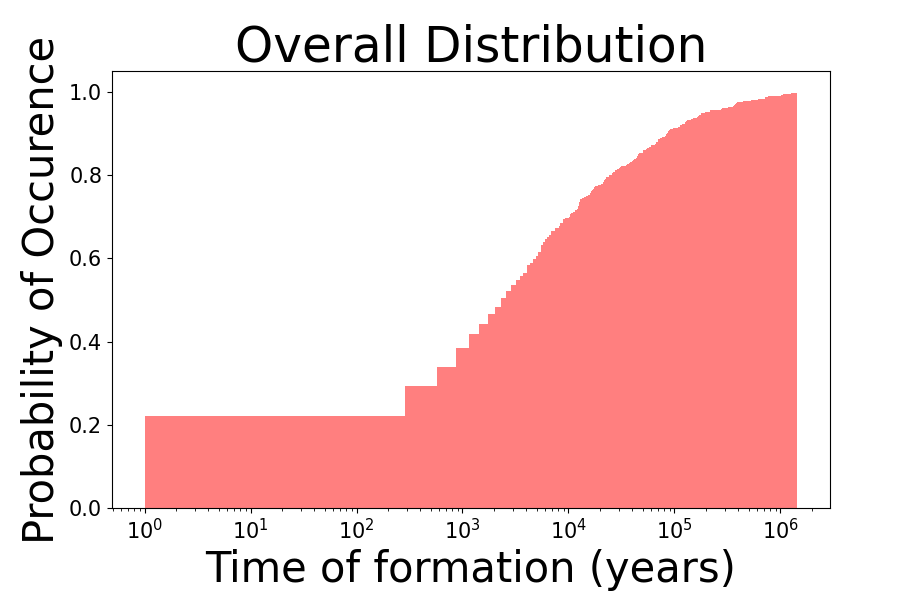}
  \end{tabular}
\end{tabular}

\caption{Cumulative distribution for the time of formation for every binary planet formed. On the left, results from simulations with two (blue), three (green), four (pink), and five (yellow) initial planets. On the right, overall cumulative distribution obtained by combining the previous data sets.}
\label{fig5}
\end{figure*}

Figures \ref{fig2}, \ref{fig3}, \ref{fig4}, and \ref{fig5} display the distributions of the key physical parameters for the formed binaries. In each figure, the left panels present the distributions for different initial numbers, including $N=2$ (top left), $N=3$ (top right), $N=4$ (bottom left), and $N=5$ (bottom right) initial planets. The right panel, on the other hand, represents the cumulative distributions obtained by considering all the formed binaries, regardless of $N$. For these four sets of simulations we also run the Kolmogorov-Smirnov test \citep[K-S test, ][]{Hodges} to check for similarities and differences in the distributions of the mass ratio, time of formation, semi-major axis and eccentricity. The K-S test was performed between pairs of samples and results are presented in Appendix A and in Table \ref{KS}.

Figure \ref{fig2} and \ref{fig3} illustrate the distributions of eccentricities and semi-major axes for binary planets that remained in a stable orbit around the star. Such parameters are relative to the orbit of the binary, considered as one single object, around the host star. Regarding eccentricity, the peak of the distribution gradually shifts from lower values ($<0.2$) for two initial planets to intermediate/high values for five planets, with the majority of binaries having eccentricities in the range of $0.2$ to $0.6$. The differences in these four distributions are highlighted by the results of the K-S test, with p-values ranging from a maximum of $10^{-2}$ for sets 2 and 3 to a minimum of $10^{-54}$ for set 1 and 4. In fact, the differences among the four distributions are particularly relevant when considering two planets in comparison with the initial configuration with $N>2$. However, the overall distribution across the four sets indicates a preference for moderately low eccentricities, with a peak at 0.1. The distributions of semi-major axes as well differ for the first four sets of simulations, showing low p-values (see Table \ref{KS}). However, we note that the majority of binaries orbit their star at distances less than 100 au and that there is a preference for inward migration of binary planets ($39\%$ for binaries located within $<50$ au) compared to outward migration ($13\%$ for binaries located at $>100$ au), considering that the initial semi-major axis was randomly chosen in the range $[50,100]$ au. One possible explanation could reside in the fact that a bound pair is an heavier than a single planet and thus tend to eject the latter.

In Figure \ref{fig4} and \ref{fig5}, we present the distribution of the last two parameters: mass ratio and time of formation, respectively. These plots consider the entire sample of binary planets formed, irrespective of their final status (bound, ejected, or collided). The mass ratio exhibits a similar behaviour for each $N$ considered (see values in Table \ref{KS}), with a preference for binaries with a mass ratio close to one. As shown in the cumulative distribution, most disfavoured binaries have mass ratios ranging from $0.1$ to $0.3$.

On the other hand, the formation time appears to be dependent on the initial number of planets, showing a significantly different distribution especially for set 1 with respect to set 2, 3, and 4. Notably, however, the binary planets formation process through gravitational capture enhanced by tides is quite rapid. After the individual planets are formed through gravitational instability \citep[$10^3-10^5$ yrs,][]{Boss} it typically takes less than $10^{3}$ years for the system to develop planet-planet pairs.

In conclusion, apart from the mass ratio distribution which is not strongly influenced by the initial number of planets, the distributions of other parameters are dependent on N. In particular, we note that the distributions that are most distant from all the others for eccentricity, semi-major axis, and formation time differences come from the initial condition of $N=2$.

\subsection{Parameters dependency on the strengths of the tides}

In sets 5, 6, and 7, we explored the variation in the amount of energy lost due to tidal interaction during close encounters. The precise nature of this term is challenging to determine due to the complex processes involved, and it can be both overestimated and underestimated. While it is certain that some form of friction occurs when young giant planets come into close proximity, the physics underlying such interactions is difficult to constrain. The gas dynamics can involve the gaseous envelopes of the planets, the circumplanetary disks, or both.

\begin{figure}
\centering
\includegraphics[width=\columnwidth]{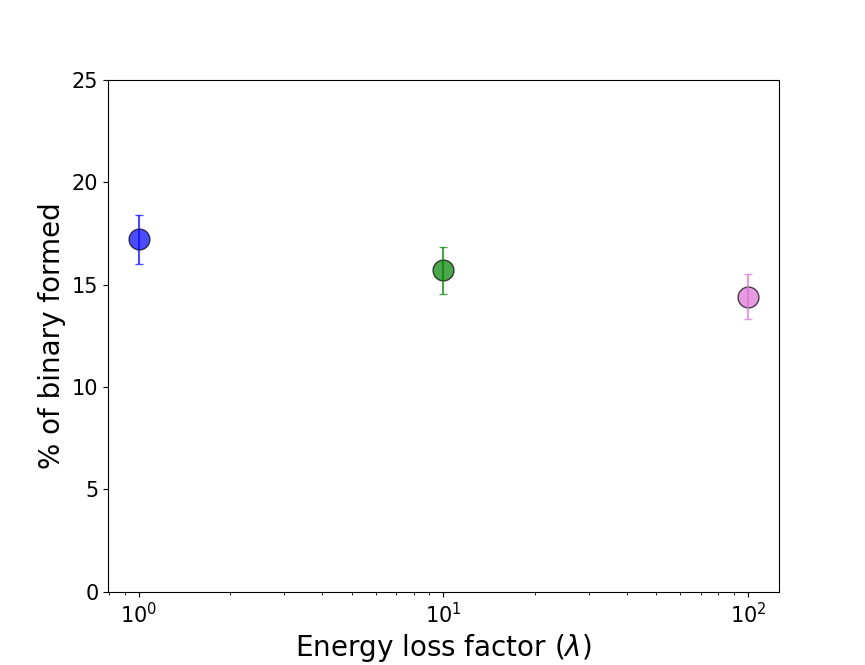} 
\caption{Percentage of binary planets formed for strong ($\lambda=1$), intermediate ($\lambda=10$) and weak ($\lambda=100$) tides.}
\label{fig6}
\end{figure}

To account for this uncertainty, we initially considered a general term for tidal dissipation in set 5 and reduced it by a factor of 10 in set 6 and 100 in set 7. As demonstrated in Figure \ref{fig6}, even if the energy loss constitutes only a fraction as small as $1/10$ or $1/100$ of the initial quantity considered, the formation of binary planets through tides proves to be an efficient process, with a formation rate of $15.7\% \pm 1.1\%$ and $14.4\% \pm 1.1\% $, respectively. As expected, the formation rate gradually decreases when less energy is lost due to tidal friction. The errors are calculated as explained in Section 3.1.

\begin{figure*}

\centering

\begin{tabular}{@{} c c c @{}}

    \includegraphics[width=6cm,height=5cm]{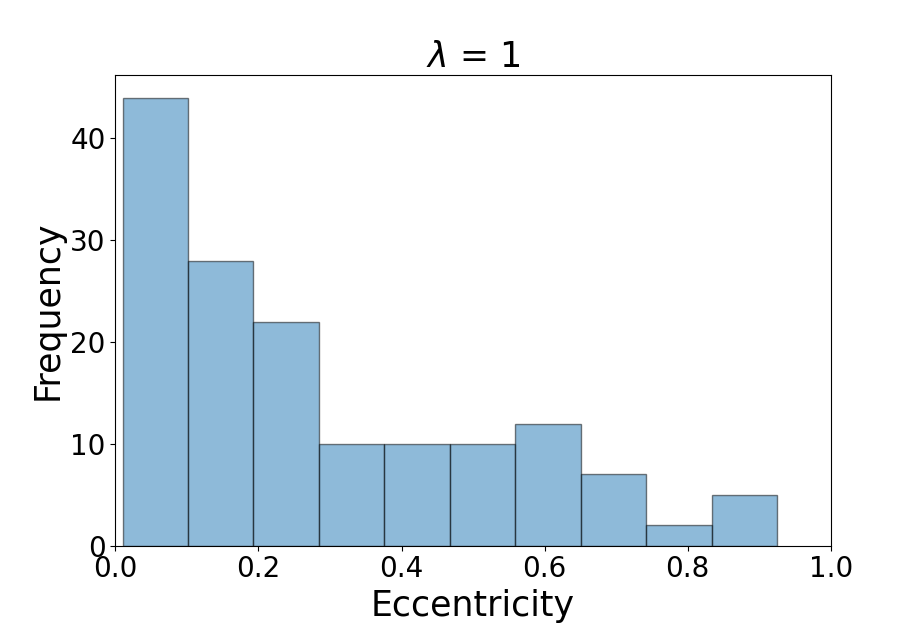} &
    \includegraphics[width=6cm,height=5cm]{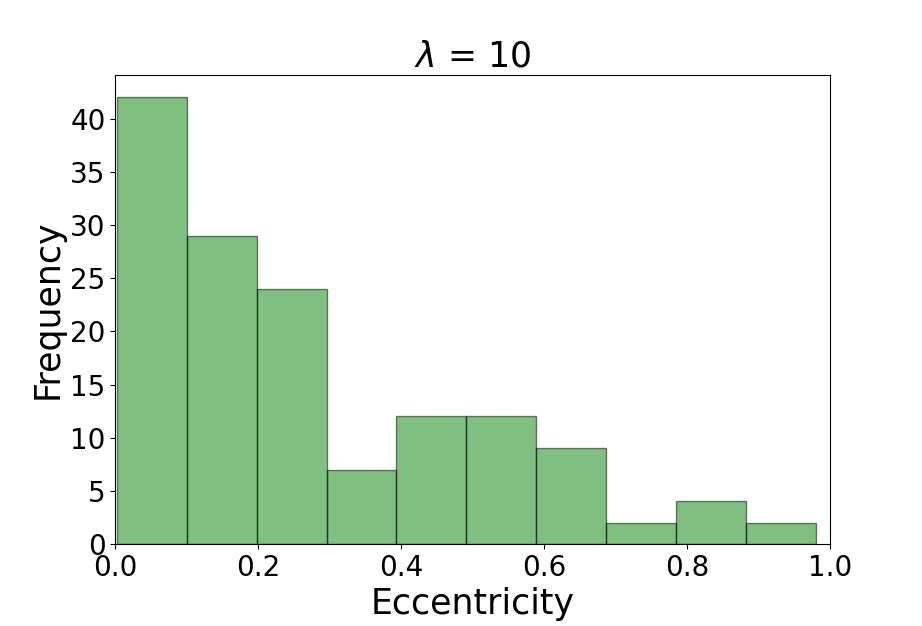} &
    \includegraphics[width=6cm,height=5cm]{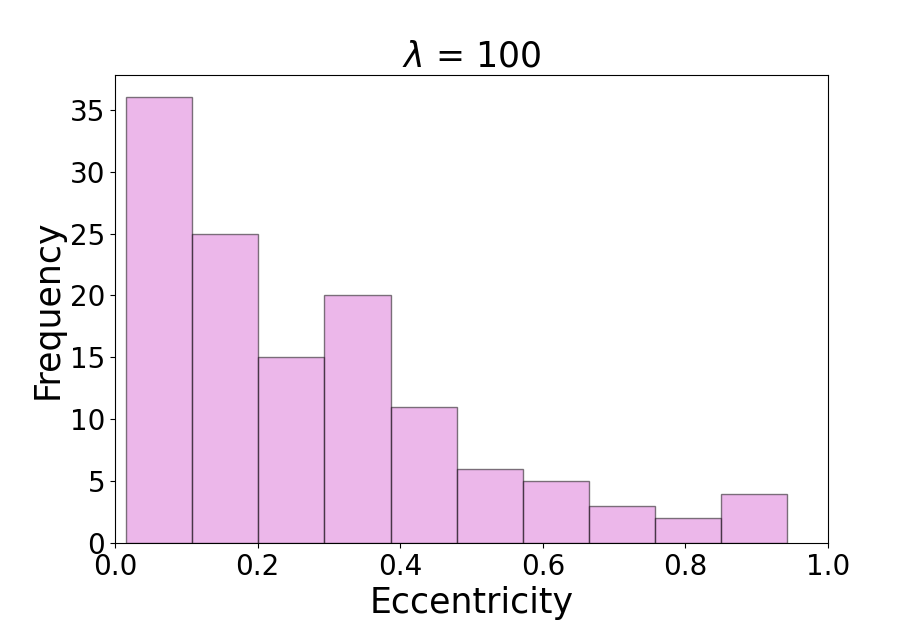} 
\end{tabular}
\caption{Eccentricity distributions for binary planets on stable orbits around the star. The eccentricity is relative to the orbit of the binary, considered as one object, around the star. From the left, results from simulations with strong ($\lambda=1$, blue), intermediate ($\lambda=10$, green) and weak ($\lambda=100$, pink) tides.}
\label{fig8}
\end{figure*}

\begin{figure*}

\centering

\begin{tabular}{@{} c c c @{}}

    \includegraphics[width=6cm,height=5cm]{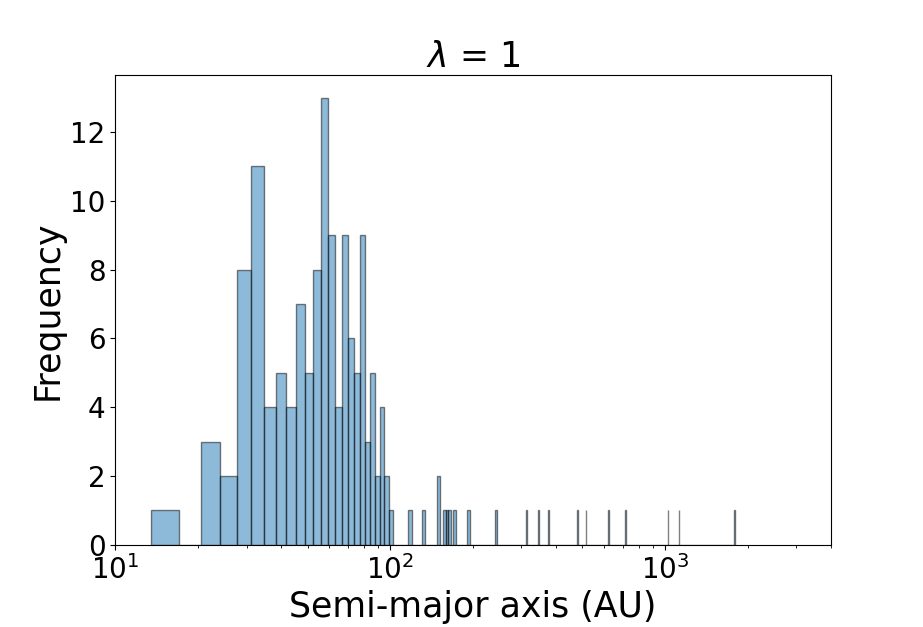} &
    \includegraphics[width=6cm,height=5cm]{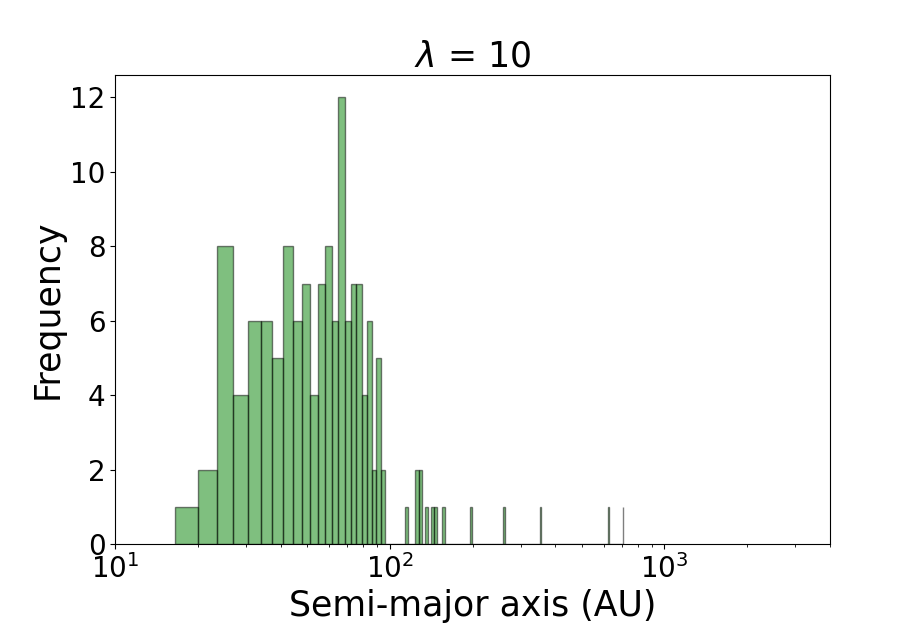} &
    \includegraphics[width=6cm,height=5cm]{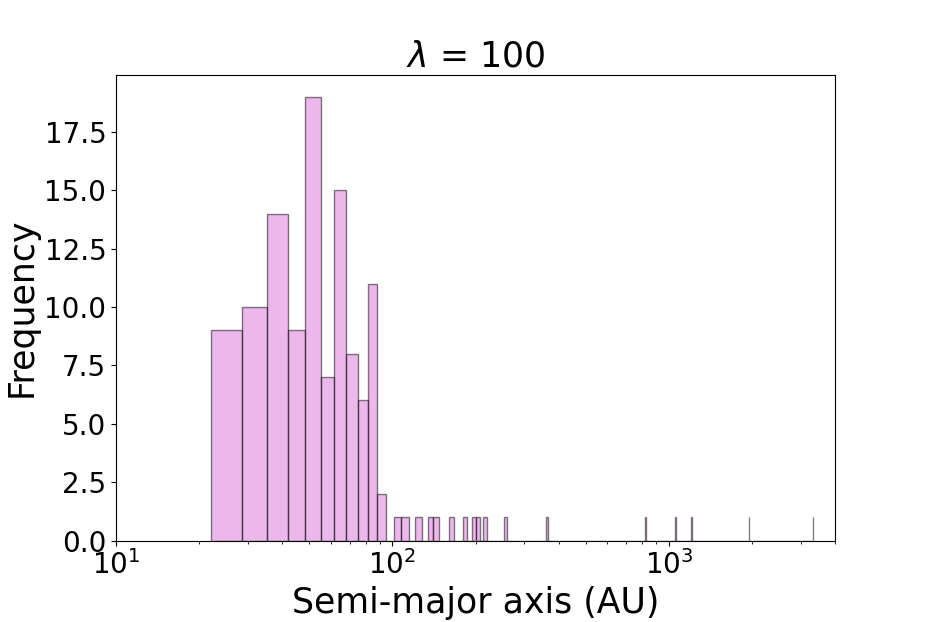} 
\end{tabular}
\caption{Semi-major axis distributions for binary planets on stable orbits around the star. The semi-major axis is relative to the orbit of the binary, considered as one object, around the star. From the left, results from simulations with strong ($\lambda=1$, blue), intermediate ($\lambda=10$, green), and weak ($\lambda=100$, pink) tides.}
\label{fig9}
\end{figure*}

\begin{figure*}

\centering

\begin{tabular}{@{} c c c @{}}

    \includegraphics[width=6cm,height=5cm]{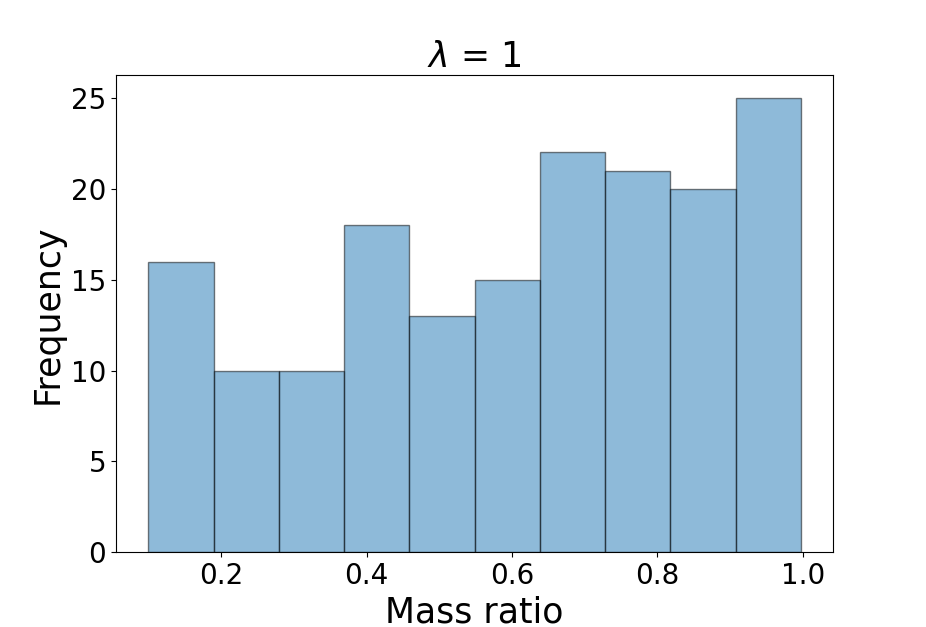} &
    \includegraphics[width=6cm,height=5cm]{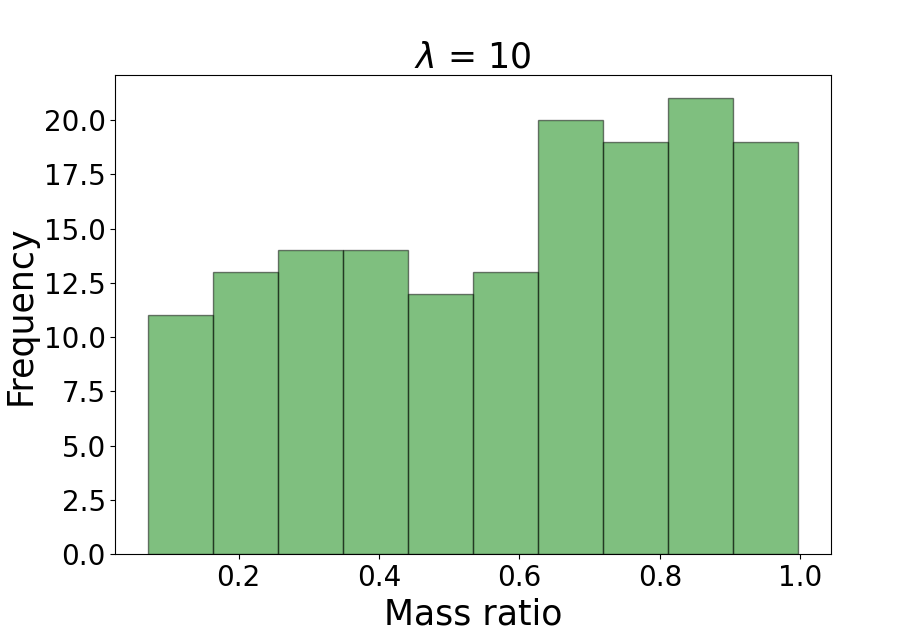} &
    \includegraphics[width=6cm,height=5cm]{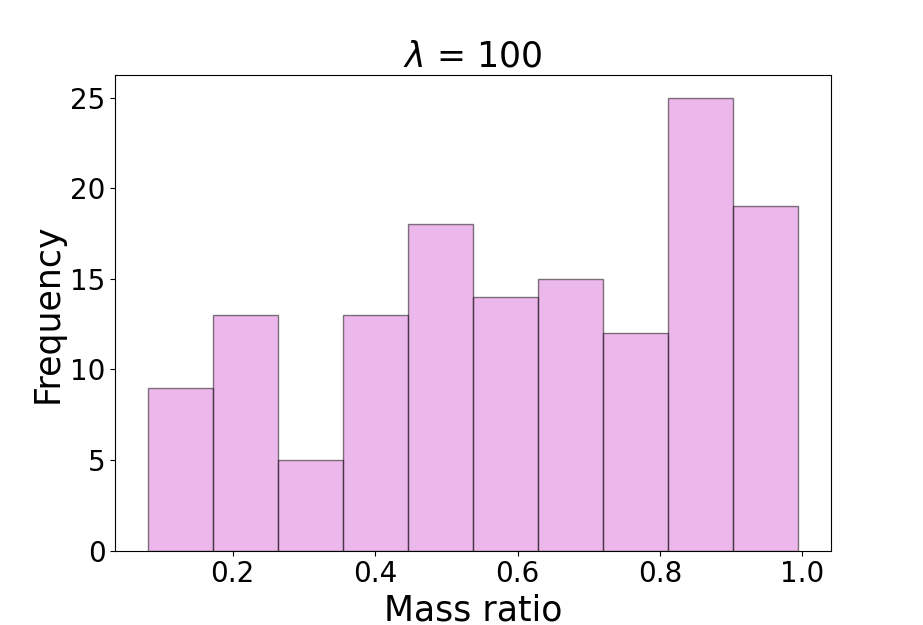} 
\end{tabular}
\caption{Mass ratio distributions for binary planets on stable orbits around the star. From the left, results from simulations with strong ($\lambda=1$, blue), intermediate ($\lambda=10$, green) and weak ($\lambda=100$, pink) tides.}
\label{fig10}
\end{figure*}

\begin{figure*}
\centering
\begin{tabular}{@{} c c c @{}}

    \includegraphics[width=6cm,height=5cm]{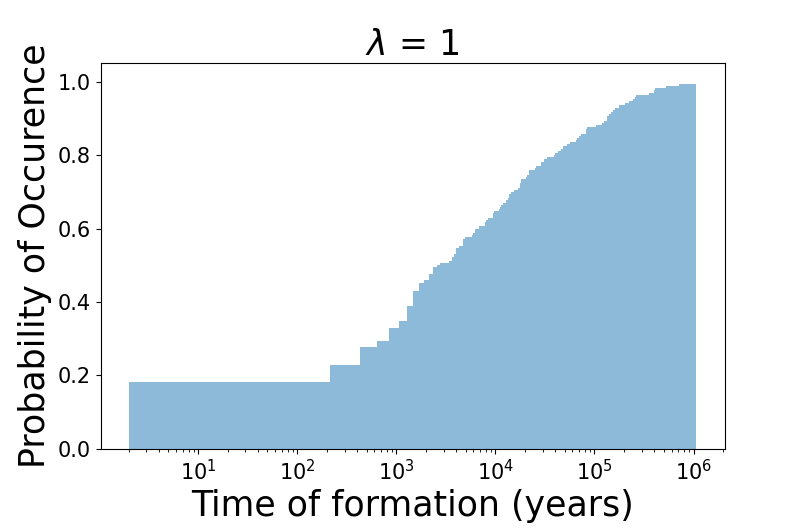} &
    \includegraphics[width=6cm,height=5cm]{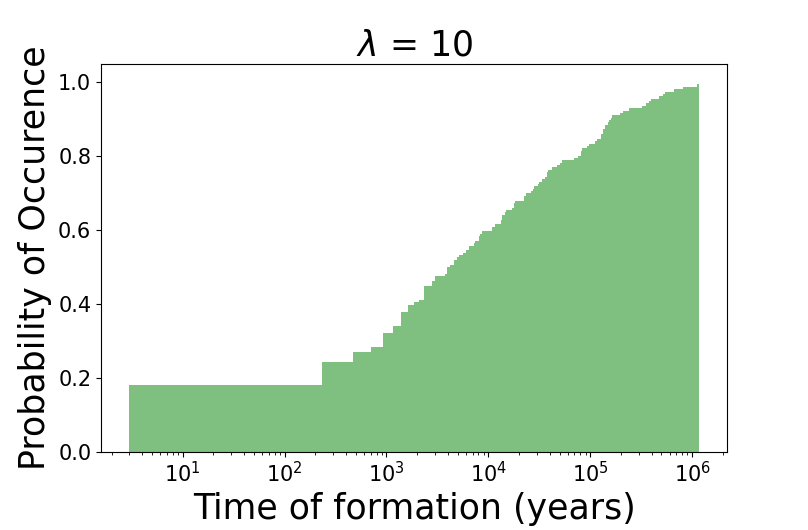} &
    \includegraphics[width=6cm,height=5cm]{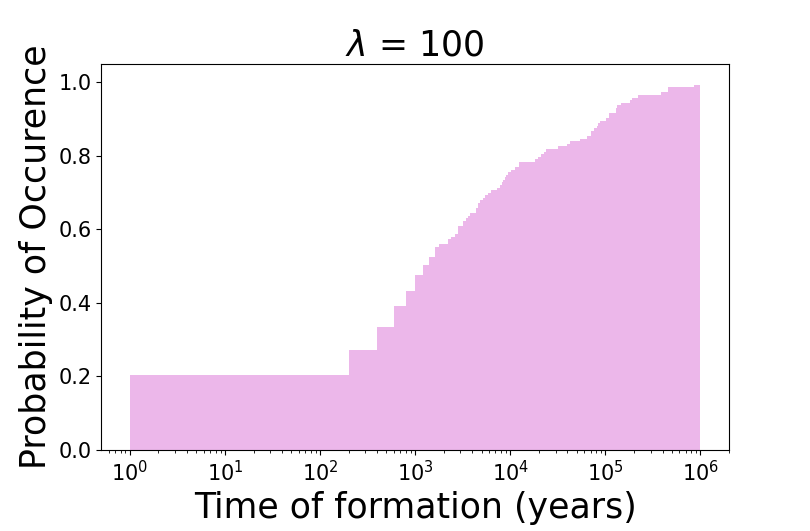} 
\end{tabular}
\caption{Time of formation cumulative distributions for binary planets on stable orbits around the star. From the left, results from simulations with strong ($\lambda=1$, blue), intermediate ($\lambda=10$, green) and weak ($\lambda=100$, pink) tides.}
\label{fig11}
\end{figure*}

In Figure \ref{fig8}, \ref{fig9}, \ref{fig10}, and \ref{fig11}, we present the distributions of eccentricity, semi-major axis, mass ratio, and time of formation for the binaries formed with $\lambda=1, 10, 100$. Similar to the plots in Section 3.1, only binaries in stable orbits around the star at 1.5 Myrs were considered for the eccentricity and semi-major axis distributions, while all formed binaries (as indicated in column seven of Table \ref{tab2}) were considered for mass ratio and time of formation distributions. We also stress that eccentricities and semi-major axes are relative to the orbit of the binary around the host star and not to the motion of the two planets around their center of mass. Similarly to the analysis with varying initial number of planets, we run the K-S test for these three samples, comparing the distributions of the key phisical parameters. The K-S test was once more performed between pairs of samples.

Both eccentricity and semi-major axis do not exhibit a strong dependence on the strength of tides, showing a similar slope at each $\lambda$ (see Table \ref{KS} for results of the K-S test and Figures \ref{fig8} and \ref{fig9}). As obtained for the first four sets of simulations, there is a preference for low eccentricities (<0.2) and semi-major axes belonging to the original injection range of $[50,100]$ au.

Similarly, the mass ratio and the formation time does not exhibit a clear dependence on $\lambda$. This last result also confirms that this formation mechanism is rapid even when tides are not that strong.

\subsection{Statistics on single planets and binaries}

\begin{figure*}

\centering

\begin{tabular}{@{} c c @{}}
  \begin{tabular}{@{} c c @{}}
    \includegraphics[width=5cm,height=4cm]{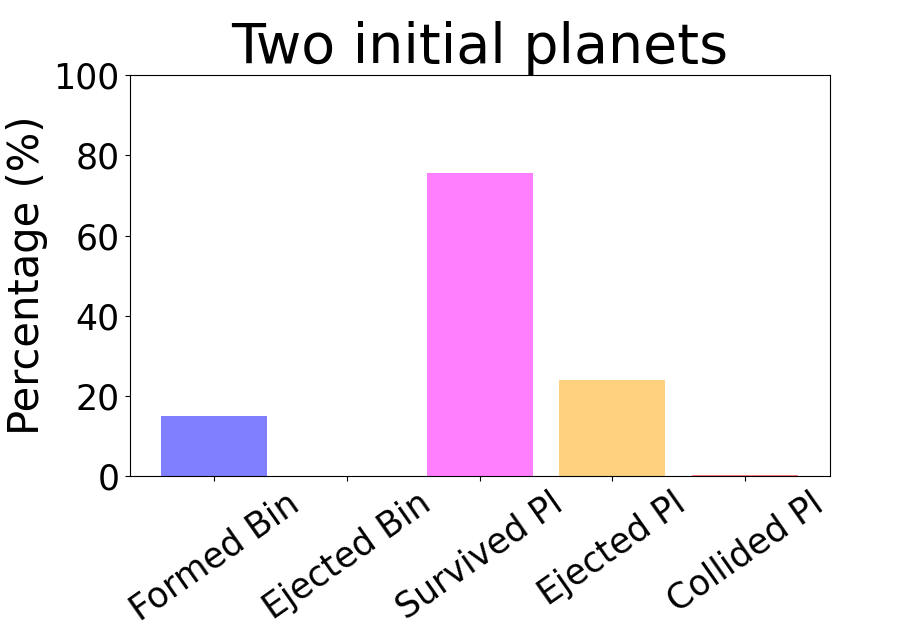} &
    \includegraphics[width=5cm,height=4cm]{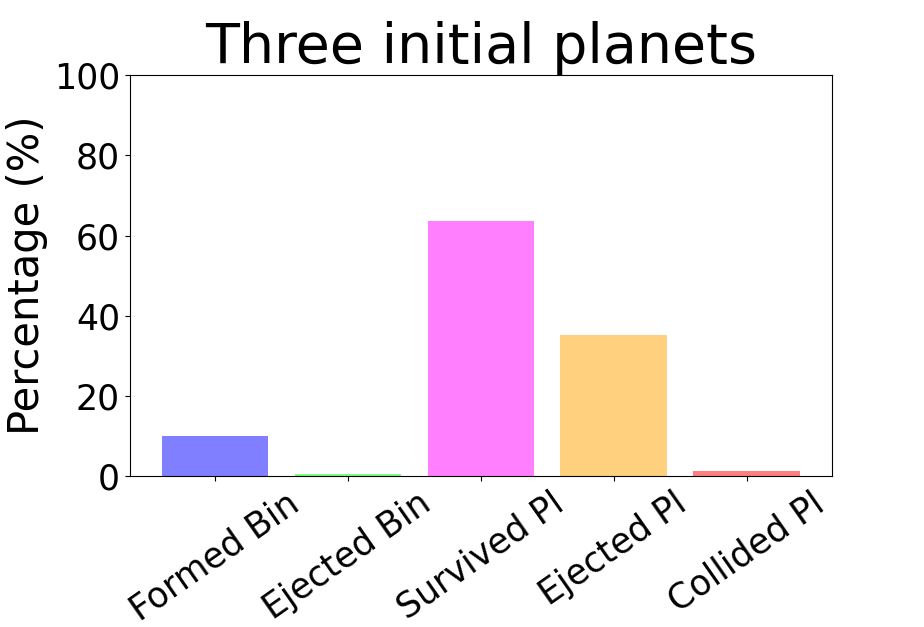} \\[2ex]
    \includegraphics[width=5cm,height=4cm]{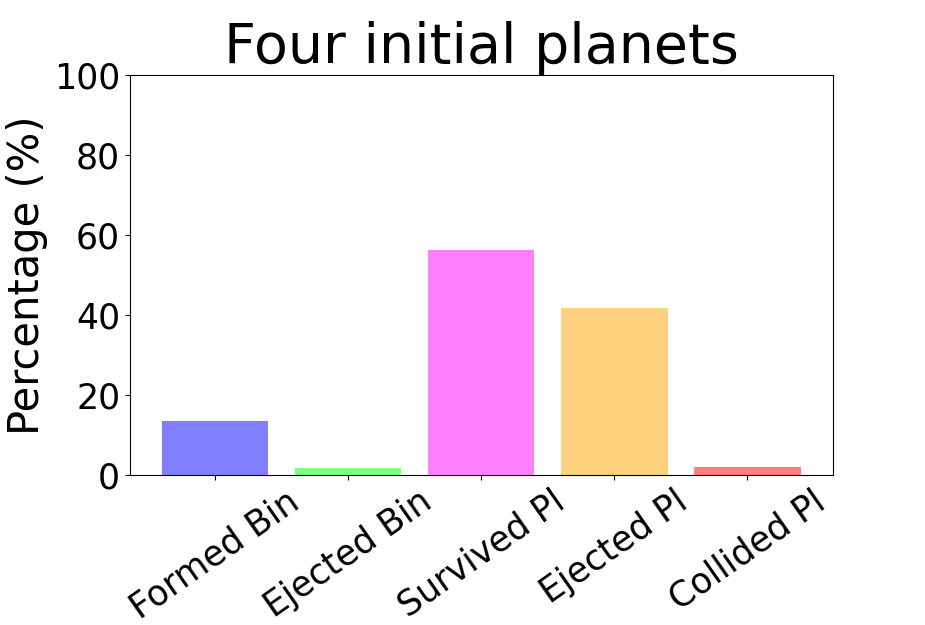} &
    \includegraphics[width=5cm,height=4cm]{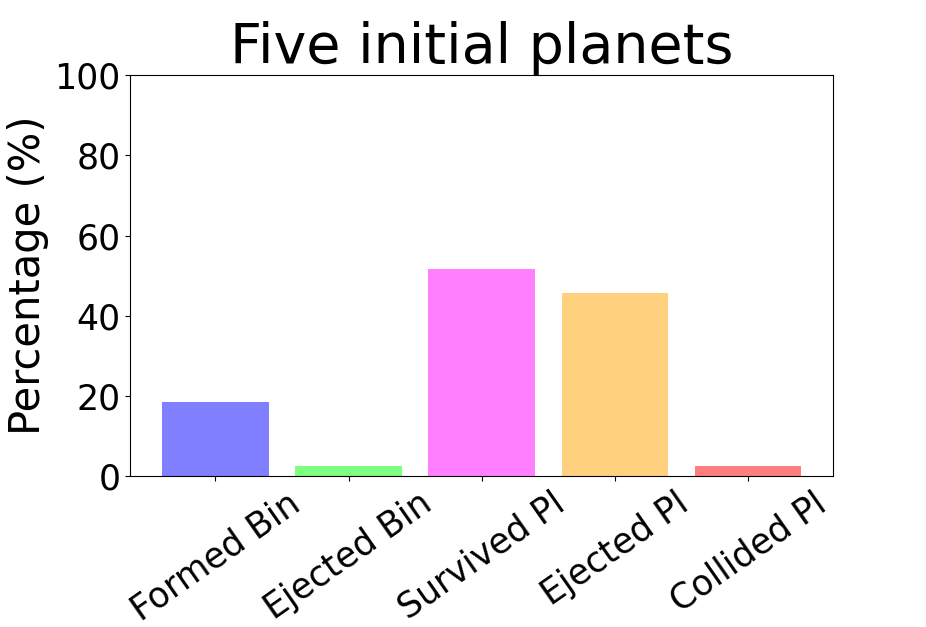}
  \end{tabular}
&
  \begin{tabular}{@{} c @{}}
    \includegraphics[width=7cm,height=6cm]{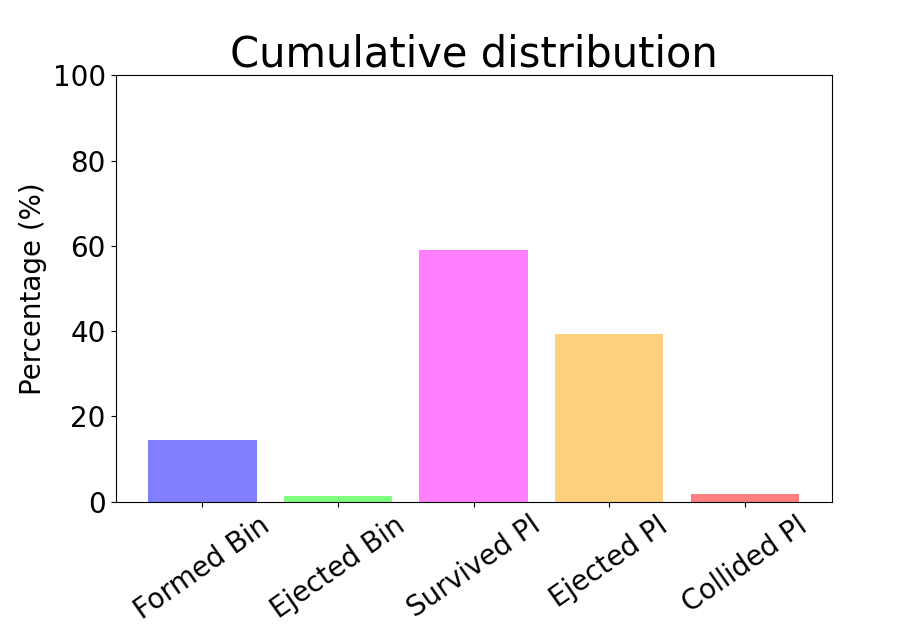}
  \end{tabular}
\end{tabular}

\caption{In each frame is shown the percentage of binary planets formed (blue) and ejected (green), planets that survived (pink), that were ejected (orange), and that collided with the central star (red). In the four images on the left are shown the results for two (top left), three (top right), four (bottom left), and five (bottom right) initial planets; on the right instead is shown the result for the cumulative distribution.}
\label{fig12}
\end{figure*}

In Figure \ref{fig12}, we present the statistics for the first four sets of simulations (left plots), as well as the cumulative results when combining them (right plot). Each panel displays the rate of systems that formed binary planets (blue) and with ejected binaries (green), the percentages of planets in stable orbits around their host star (pink), of ejected planets (orange), and of planets that collided with the star (red). The statistics for collided binaries are not shown in the plots due to the small number of cases (rates of $0.1\%$ for set 2 and $0.02\%$ for the global statistics). However, they were taken into account in the statistics for planets that collided with the central star, along with the counts for survived and ejected planets, where each binary counts as two planets.

The results show that systems with two initial planets can form binaries at higher rates than systems with three or four initial planets. This indicates that frequent dynamical perturbations caused by a crowded system dominate over tidal dissipation and could inhibit the formation of binaries. However, with the exception of $N=2$ which constitutes the less chaotic configuration, the rate of systems able to form binaries increases significantly with increasing $N$.   

The percentage of systems with ejected binaries, instead, steadily increases from $N=2$ with a $0\%$ rate to a maximum of $2.6\%$ for five initial planets. This is true also for ejected planets, for which the rate varies from $24\%$ for $N=2$ to $45.6\%$ for $N=5$. The percentage of planets that collided with the central star is generally lower but still correlated with $N$. 

The cumulative results for sets 1, 2, 3, and 4 reveal that the rate of systems able to form binary planets is $14.3\%$. However, a $2.42\%$ of such systems will lose the planet-planet pair due to ejection ($2.4\%$) or collision with the central star ($0.02\%$). This indicates that $11.9\%$ of the systems can host bound binary planets. Additionally, we obtain constraints on the population of free-floating sub-stellar objects, with a cumulative rate of $39.2\%$.

\begin{figure*}
\centering
\begin{tabular}{@{} c c c @{}}

    \includegraphics[width=6cm,height=5cm]{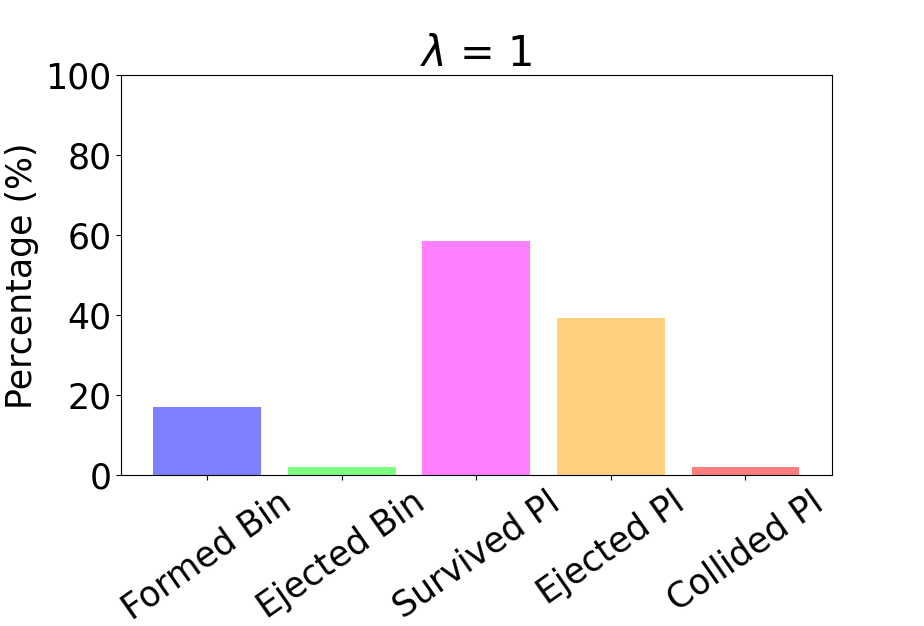} &
    \includegraphics[width=6cm,height=5cm]{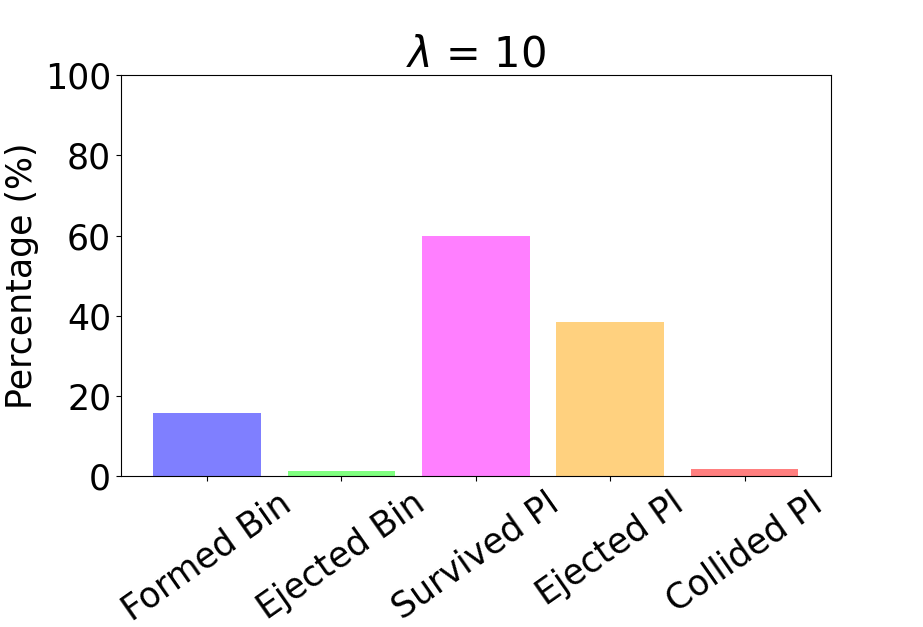} &
    \includegraphics[width=6cm,height=5cm]{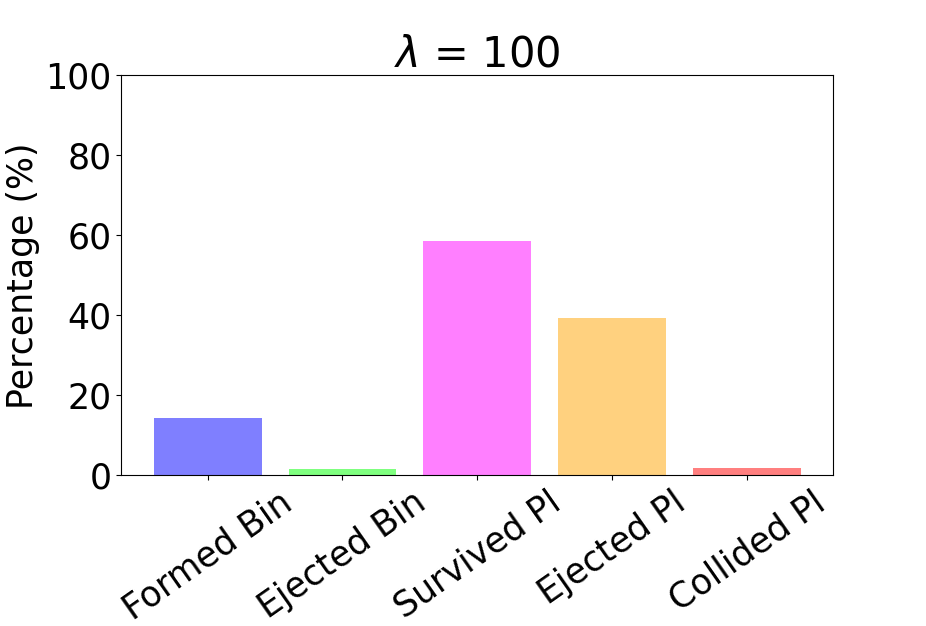} 
\end{tabular}
\caption{In each frame is shown the percentage of binary planets formed (blue) and ejected (green), planets that survived (pink), that were ejected (orange), and that collided with the central star (red). From left to right, results for strong ($\lambda=1$), intermediate ($\lambda=10$), and weak ($\lambda=100$) tides.}
\label{fig13}
\end{figure*}

In Figure \ref{fig13}, the statistics for bound, ejected, and collided single and binary planets are presented while varying the intensity of tides in sets 5, 6, and 7. As observed in the previous Section, the parameter that is most affected is the rate of systems forming binary planets, with a gradual decrease of the latter with increasing $\lambda$ ($17.2\%$ for $\lambda=1$, $15.7 \%$ for $\lambda=10$ and $14.4\%$ for $\lambda=100$).

\section{Discussion}
As demonstrated by the results presented in Section 3, binary planets can be common in planetary systems where gravitational instability is at play. With a rate of $11.9\%$, one out of every ten systems that form planets via GI should host a binary planet. However, despite a few candidates such as the previously mentioned DH Tau Bb and Kepler 1625 b-i, observational surveys struggle to confirm the existence of binary planets. This can be attributed to multiple factors.

Firstly, the physics of tides is influenced by various phenomena. The specific factor considered in our simulations is a general one and may not fully represent the entire zoo of planets under consideration. Particularly, if the tidal interaction is weak, the formation of binary planets may be significantly reduced.

Secondly, although we know that circularization and synchronization of orbits are common for paired gaseous objects, as in the case of a Hot Jupiter close to its star \citep[e.g.,][]{Rasio,Jackson}, there is a possibility that a percentage of formed binaries approach each other at high velocities, preventing them from dissipating energy efficiently and leading to collisions.

Thirdly, gravitational instability is expected to form planets at large separations from the star where radial velocities and transits are ineffective for detection. Direct Imaging \citep{Currie3} is the most suitable technique to detect this population of substellar objects. However, recent surveys have revealed that giant planets on wide-orbits are not as abundant as previously anticipated \citep{Vigan2,Nielsen1}, which may indicate that protostellar systems rarely satisfy the conditions for planet formation via GI \citep{Rice4}. Consequently, even if binary planets might be relatively common, the scarcely dense population of gravitational instability planets may explain the very small number of candidates.

Lastly, we anticipate that the separations between pairs of substellar objects are relatively small (in our simulations, we stopped at 0.1 au) \citep{Ochiai}. Current instruments are not capable of physically resolving two-point spread functions that are located in such close proximity. For instance, an instrument like the Integral Field Spectrograph of SPHERE (IFS/SPHERE) \citep{Claudi, Mesa}, with a pixel scale of 7.46 mas/pixel, achieves a resolving power of 0.05" in the H-band. For a system located at 10 parsecs, this corresponds to approximately 0.5 au, which is five times larger than the separation considered in our simulations.
On the other hand, the techniques aimed at the detection
of very close companions around directly imaged planets, such as radial velocity monitoring and transit searches, are still in an exploratory phase  
\citep{Vanderburg2021,Ruffio2023,Biller2021} although there are prospects for significantly improving detection capabilities with new instrumentation 
\citep{Lazzoni2022,vigan2023}.

The microlensing technique has also the potential to detect
pairs of massive planets or brown dwarfs. A few such cases have been reported
\citep{Choi2013,Han2017,Albrow2018} as well as some ambiguous cases \citep{Miyazaki2018} or candidate star-planet-moon systems \citep{Hwang}. These cases may resemble the ejected binaries
seen in a small percentage in our simulations, although they may have formed as isolated pairs through turbulent fragmentation.
Unfortunately, we are not aware of robust statistical inferences on the frequency of such configurations in microlensing surveys, leaving the discussion on the origin of such objects rather speculative.

The only binary candidate observed through direct imaging is DH Tau Bb \citep{Lazzoni2}. Given its large orbital distance of 300 au and slow revolution period around the star, the determination of the semi-major axis and eccentricity of DH Tau Bb is currently challenging. However, based on the results presented in Section 3.1, we can assume that the eccentricity ranges from low ($0.1-0.2$) to intermediate values ($0.4-0.5$). If the detected position corresponds to the apastron passage of the binary, this could move the semi-major axis as close as 200 au. Therefore, assuming the most likely values for the eccentricity, DH Tau Bb would be situated towards the right end of the distribution shown in Figure \ref{fig3}.

The mass ratio of 0.1 obtained for DH Tau Bb is also slightly anomalous since the distribution peaks at 1. However, there is still a consistent population of binaries with lower mass ratios as shown by the right panel of Figure \ref{fig4}.

What significantly deviates from the assumptions made in this analysis is the separation between the two planets. As reported in \cite{Lazzoni2}, DH Tau Bb orbits at a mutual distance of 10 au. As explained earlier in the paper, we would typically expect binary planets to be much closer due to mutual energy loss resulting from tidal friction, which eventually leads to the circularization of the orbit and the synchronization of rotation. The case of DH Tau Bb could indicate that circularization can occur earlier, allowing binary planets to exist with wider separations. Another possibility is that the orbit is still contracting. The system is very young, with an estimated age of 1.4 Myrs \citep{Feiden}, suggesting that the binary might have recently formed. However, it is worth noting that the binary formation process is usually rapid, and this second explanation may not be applicable.

The recent result of the moderately high binary fraction of free-floating planetary-mass objects in Orion \citep{Pearson} and the comparison with the multiplicity of brown dwarfs is also relevant for our discussion. 
While the separations of the observed planets are much larger than the binary planets generated in our simulations, their finding points to
the presence of specific mechanism(s) generating binary planets, not occurring, or occurring with lower efficiency, for brown dwarfs and low mass stars. 
We also note that the environment of Trapezium cluster considered by \citet{Pearson} has a much larger stellar density than the nearby star-forming regions and young associations that represent the typical targets of direct imaging surveys, making dynamical interactions more frequent. 

We also observe that binary planets are almost as abundant among planets formed through gravitational instability as to those formed through core accretion. In fact, \cite{Ochiai} found a formation rate of $10\%$ for core accretion models, and in our work we found a rate of $11.9\%$. This small discrepancy can be attributed to several factors. GI planets, in fact, not only exert a stronger gravitational influence on their surroundings due to their higher masses, but they also orbit at wider distances, weakening the influence of the central star and facilitating the formation of binary planets at larger semi-major axes.




\section{Conclusions}

In this study, we investigated the formation of binary planets through gravitational capture enhanced by tidal friction. This work serves as a continuation of the research conducted by \cite{Ochiai}, who focused on core accretion Jupiter-like planets in the range of 1 to 20 au. In contrast, we explored gravitational instability planets, which are expected to be more massive (1-15 M\textsubscript{Jup}) and located on wider orbits (50-100 au). The determination of tidal interactions between giant planets is a complex task. Therefore, we began with a general term proposed by \cite{Ochiai} and subsequently considered various fractions of it. Moreover, we investigated multiple scenarios varying the number of initial planets from a minimum of two to a maximum of five.

The main findings can be summarized as follows:
\begin{compactitem}
\item in the scenario where planets experience repeated close encounters characterized by tidal friction, systems can form binary planets at a rate of $14.3\%$, with $2.4\%$ losing the planet-planet pair due to ejection and a $0.02\%$ due to collision with the central star. $11.9\%$ of the systems can then host a bound binary on a stable orbit. This indicates that one out of every ten systems may host binary planets, similarly to what was obtained by \citep{Ochiai};
\item the number of initial planets, $N$, influences the formation rate of binaries. $N=2$ is particularly suitable for the formation of binaries with a rate of $15.1\%$. More planets in the system imply also more dynamical perturbations that conflict with the formation of planet-planet pairs. However, the rate increases steadily from $10.1\%$ for $N=3$ to $18.6\%$ for $N=5$;
\item the strength of tides plays a significant role in the formation of binary planets. We explored three scenarios: strong tides ($\lambda=1$), intermediate tides ($\lambda=10$), and weak tides ($\lambda=100$). It was found that there is a gradual decrease in the formation rate of binary planets correlated with increasing values of $\lambda$;
\item the formation of binary planets through this mechanism is extremely rapid. The majority of planet-planet pairs are formed within $10^3$ years after gravitational instability has successfully formed single planets in the circumstellar disk;
\item most of the formed binaries, considered as single objects orbiting around the host star, exhibit low to moderate eccentricities and do not display significant migration. However, inward migration is slightly favored over outward drift.
\end{compactitem}

Overall, these findings shed light on the formation of binary planets and provide valuable insights into their characteristics and prevalence in planetary systems.

\section*{Acknowledgements}
C.L. and A.Z. acknowledge support from ANID -- Millennium Science Initiative Program -- Center Code NCN2021\_080.  K.R. is grateful for support from UK STFC via consolidated grant ST/V000594/1.
SD acknowledges support by the PRIN-INAF 2019 "Planetary systems at young ages (PLATEA).
\section*{Data Availability}
There is no new data associated with this article.



\bibliographystyle{mnras}
\bibliography{bibliography} 



\appendix

\section{Kolmogorov-Smirnov tests}
We run the Kolmogorov-Smirnov test \citep[K-S test][]{Hodges} to check if the different samples are following similar distributions. For each parameter, nominally mass, time of formation, eccentricity, and semimajor axis, we performed the two-sample K-S test combining data sets 1 to 4 and separately 5 to 7. For each combination, the p-value is given in Table \ref{KS}.

\begin{table*}
	\centering
	\caption{}
	\label{KS}
	\begin{tabular}{cccccccccc} 
		\hline
		 & Set 1-2 & Set 1-3 & Set 1-4 & Set 2-3 & Set 2-4 & Set 3-4 & Set 5-6 & Set 5-7 & Set 6-7\\
		\hline
		Mass & 0.56 & 0.41 & 0.11 & 0.70 & 0.02 & 0.14 & 0.48 & 0.90 & 0.38\\
		Time Formation & 3.73e-16 & 1.01e-25 & 2.09e-38 & 0.015 & 4.47e-05 & 0.013 & 0.58 & 0.15 & 0.023\\ 
            Semimajor axis & 9.43e-23 & 4.94e-30 & 5.17e-31 & 4.95e-06 & 1.58e-10 & 3.18e-4 & 0.078 & 0.119 & 0.718\\
            Eccentricity & 2.09e-19 & 1.87e-36 & 1.13e-54 & 8.71e-03 & 3.14e-07 & 0.074 & 0.982 & 0.557 & 0.481\\
		\hline
	\end{tabular}
\end{table*}


\bsp	
\label{lastpage}
\end{document}